\documentclass[a4paper,fleqn,usenatbib]{mnras}

\usepackage{newtxtext,newtxmath}

\usepackage[T1]{fontenc}
\usepackage{ae,aecompl}

\usepackage{graphicx}	
\usepackage{amsmath}	
\usepackage{mathtools}	
\usepackage{array}	         
\usepackage{ragged2e}    
\usepackage{sidecap}                                         
\sidecaptionvpos{figure}{c}                                  
\usepackage[font=small, labelfont=bf]{caption}   
\usepackage{multirow}        				 
\usepackage{floatrow}                                         
\usepackage{arydshln}					 
\usepackage{bigstrut}					 
\usepackage{float}						
\usepackage[export]{adjustbox}                          

\graphicspath{ {./Figures/} }               
\numberwithin{equation}{section}    

\newcommand{\Msun}{{\,M}$_\odot$}
\newcommand{\Lsun}{{\,L}$_\odot$}
\newcommand{\Dsun}{{\,D}$_\odot$}

\newcommand{\kms}{{\,\rm {km s$^{-1}$}}}

\usepackage{longtable}  			
\usepackage{supertabular}
\usepackage{natbib}                          
\usepackage[autostyle]{csquotes}
\usepackage{subfigure}                     
\usepackage{graphicx}

\title[Properties of M31's Lower NW Stream ]{Properties of the Lower Segment of M31's North West Stream}

\author[J. Preston et al]   
{Janet Preston,$^{1*}$
\ Denis Erkal,$^1$
\ Michelle L.M. Collins,$^1$
Rodrigo Ibata,$^2$ 
\ R. Michael Rich$^3$
\\
\\
$^1$Department of Physics, University of Surrey, Guildford, GU2 7XH, Surrey, UK. \thanks{j.preston@surrey.ac.uk} \\
$^2$Observatoire de Strasbourg, 11, rue de l'Universit\'{e}, F-67000, Strasbourg \\
$^3$Department of Physics and Astronomy, UCLA, 430 Portola Plaza, Box 951547, Los Angeles, CA 90095-1547, USA \\
}

\date{Accepted XXX. Received YYY; in original form ZZZ}

\pubyear{2023}

\begin{document}
\label{firstpage}
\pagerange{\pageref{firstpage}--\pageref{lastpage}}
\maketitle

\begin{abstract}
We present a kinematic and spectroscopic analysis of 40 red giant branch stars, in 9 fields, exquisitely delineating the lower segment of the North West Stream (NW-K2), which extends for $\sim$80 kpc from the centre of the Andromeda galaxy.   We measure the stream's systemic velocity as -439.3$^{+4.1}_{-3.8}$ \kms{} with a velocity dispersion = 16.4$^{+5.6}_{-3.8}$ \kms{} that is in keeping with its progenitor being a dwarf galaxy. We find no detectable velocity gradient along the stream. We determine $-$1.3$\pm$0.1 $\le$ <[Fe/H]$_{\rm spec}$> $\le$ $-$1.2$\pm$0.8 but find no metallicity gradient along the stream.  We are able to plausibly associate NW-K2 with the globular clusters PandAS-04, PandAS-09, PAndAS-10, PAndAS-11, PandAS-12 but not with PandAS-13 or PandAS-15 which we find to be superimposed on the stream but not kinematically associated with it. 
\end{abstract} 

\begin{keywords}
galaxies: dwarf -- galaxies: interactions -- Local Group
\end{keywords}

\section{Introduction} \label{Introduction}
\graphicspath{ {Figures/} }    

The Andromeda Galaxy (M31) is host to numerous stellar streams (\citealt{RefWorks:143, RefWorks:176, RefWorks:38, RefWorks:82}, \citealt{RefWorks:47}, \citealt{RefWorks:107}, \citealt{RefWorks:449}).  These spectacular structures are the ghostly remnants of dwarf galaxies and globular clusters (GCs) accreted by M31 over billions of years.  They provide tangible evidence for the $\Lambda$CDM paradigm that larger galaxies grow by devouring smaller ones (\citealt{RefWorks:513}, \citealt{RefWorks:514}, \citealt{RefWorks:515}) and enable us to explore the gravitational potentials required to produce them (e.g. \citealt{RefWorks:176}, \citealt{RefWorks:10},  \citealt{RefWorks:421},  \citealt{RefWorks:245}, \citealt{RefWorks:422},  \citealt{RefWorks:82}).

The North West (NW) stream in the outer halo of M31 has an estimated length of $\sim$200 kpc and comprises two segments: an upper segment, labelled K1 (hereafter NW-K1) in Figure 12 of \cite{RefWorks:449}; and a lower segment labelled K2 (hereafter NW-K2).  NW-K1 was discovered by \cite{RefWorks:18} using data obtained from the 3.6 m Canada-France-Hawaii Telescope (CFHT) for the Pan-Andromeda Archaeological Survey (PAndAS, \citealt{RefWorks:58}). They found this segment of the NW stream to be $\sim$3$^{\circ}$ in length at a distance of $\sim$50-80 kpc from the centre of M31. Later work by \cite{RefWorks:449} found NW-K1 to have a luminosity of \textit{M$_V$} = $-$10.5 $\pm$ 0.5 and a stellar mass of \textit{M$_*$} = 9.4 $\times$ 10$^{5}${\Msun}.

Two years earlier, NW-K2 had been discovered by \cite{RefWorks:58} who determined that it had a projected length of $\sim$6$^{\circ}$ and lay 50 $\le$ \textit{R$_{\rm{proj}}$} kpc $\le$ 120 from the centre of M31. \cite{RefWorks:449} found NW-K2 to have a luminosity of \textit{M$_V$} = $-$12.3 $\pm$ 0.5 and a stellar mass of \textit{M$_*$} = 8.5 $\times$ 10$^{6}${\Msun} and noted that it was as bright as the NGC147 stream (\textit{M$_V$} = $-$12.2, \citealt{RefWorks:449} ) and Andromeda II (\textit{M$_V$} = $-$12.6 $\pm$ 0.2, \citealt{RefWorks:206}).  

Work by \cite{RefWorks:72}, \cite{RefWorks:232} and \cite{RefWorks:233} found NW-K2 was co-located with the GCs PAndAS-04, PAndAS-09, PAndAS-10, PAndAS-11, PAndAS-12, PAndAS-13 and PAndAS-15. Subsequently, \cite{RefWorks:236} found spatial and kinematic associations between the stream and PAndAS-04, PAndAS-09, PAndAS-10, PAndAS-11, PAndAS-12 and PAndAS-13 as well as detecting that the radial velocities of the GCs became increasingly negative the nearer a GC was to the centre of M31. However, they found that PAndAS-15, the GC closest to M31, did not follow this trend and concluded that it was unlikely to be associated with NW-K2 despite seeming to lie directly on top of it.  Analysis by \cite{M:992} determined metallicities for three of the GCs: PAndAS-04 [Fe/H] = $-$2.07 $\pm$ 0.2, PAndAS-09 [Fe/H] = $-$1.56 $\pm$ 0.2 and PAndAS-11 [Fe/H] = $-$2.16 $\pm$ 0.2. Work by \cite{RefWorks:516} determined the specific frequency (which connects the number of GCs hosted by a galaxy to its total luminosity) of the NW-K2 GCs to be $\sim$70-85, which they noted was much higher than values found for other dwarf galaxy progenitors of streams around M31. This, they concluded, could be due to a much higher luminosity galaxy, probably totally destroyed with any residual debris lying within M31's inner halo, being the progenitor of NW-K2. 

Despite NW-K1 and NW-K2 being discovered separately, their morphology, as understood at the time, led \cite{RefWorks:18} to conclude that they were part of a single structure wrapped around M31.  Corroborating evidence was reported by \cite{RefWorks:82} who detected similar metallicities ($-$1.7 <[Fe/H] < $-$1.1) in both segments.   Meanwhile, noting similar density variations and gaps in the two segments \cite{RefWorks:76} modelled the stream as a single structure finding it to be $\sim$10 Gyrs old and $\sim$5 kpc wide.  They estimated the luminosity of NW-K2 to be \mbox{7.4 $\times$ 10$^5$ \Lsun} and noted that it was nearly intact while NW-K1 had some significant gaps possibly due to interactions with dark matter sub-haloes (\citealt{M:1046}).

On discovering the dwarf spheroidal (dSph) galaxy Andromeda XXVII (And XXVII) to be co-located with NW-K1, \cite{RefWorks:18} postulated that it was likely to be the progenitor of the whole of the NW Stream.  This view was sustained by \cite{RefWorks:76}, based on the number of GCs associated with NW-K2 being consistent with the progenitor being a dwarf galaxy, and by \cite{RefWorks:313} who noted that NW-K2’s progenitor would need to have a stellar mass $\sim$10$^{6}$$-$10$^{8}${\Msun} and a minimum \textit{r$_h$} $\ge$ 30 pc, which did not rule out And XXVII as being a plausible progenitor for both it and NW-K1. 

However, \cite{RefWorks:586} cast doubt on the NW stream being a single structure with their findings of a velocity gradient of \mbox{1.7 $\pm$ 0.3 \kms degree$^{-1}$} along NW-K1 that could be indicative of an infall trajectory onto M31. When considered in conjunction with the detection of a similar velocity gradient along the NW-K2 GCs, also most likely on an infall trajectory onto M31, by \cite{RefWorks:232, RefWorks:236} and findings that both NW-K1 (heliocentric distance = 827 $\pm$ 47 kpc, \citealt{RefWorks:18}, \citealt{RefWorks:586}) and NW-K2 (distance modulus $\sim$24.63±0.19, \citealt{RefWorks:405}) lie behind M31, \citeauthor{RefWorks:586} concluded that this could indicate that the two segments were not parts of a single structure. 

In this work we aim to analyse the kinematic and spectroscopic properties of NW-K2 such that we can compare them with those of NW-K1 and see if there is any association between the two streams.   We also aim to confirm, or otherwise, the association of the GCs PAndAS-04, PAndAS-09, PAndAS-10, PAndAS-11, PAndAS-12, PAndAS-13 and PAndAS-15 with NW-K2. This information will then further inform modelling of the trajectories of the two streams to more fully understand the nature of the enigmatic NW stream (\citealt{M:1148}).

The paper is structured as follows: Sections \ref{Observations} describe our observational data and approach to data reduction. Section \ref{Analysing membership of NW-K2} includes determination of candidate NW-K2 stars and our kinematic and spectroscopic analyses. We present a discussion of our results in Section  \ref{Discussion} and our conclusions in Section \ref{Conclusions}.

\begin{table*}
	\centering
	\setlength\extrarowheight{2pt}
	\caption[Properties for NW-K2 observed fields]
	{Properties for the observed fields along NW-K2, including: mask name; date observations were made; observing PI; right ascension and declination of the centre of the field, the projected radius of the mask centre from the centre of M31 and the number of confirmed members in the stellar populations for each field. The $\alpha$ and $\delta$ for the centre of each mask are determined by taking the mean of the coordinates for all stars on the mask.  The masks are listed in order of increasing distance from M31.  $^{(a)}$ are stars that probably belong to M31 or MW stellar populations, but as there were no candidate NW-K2 stars on this mask, these data were not analysed. $^{(b)}$ are stars in the NW-K2 velocity range that do not lie on its RGB.}		
	\label{K2_table:1}
	\begin{tabular}{lclcccccccc} 
		\hline
		\multirow{2}{*} {Mask name} & 
		\multirow{2}{*} {Date} &
		\multirow{2}{*} {PI} &
		\multirow{2}{*} {$\alpha_{\rm J2000}$} & 
		\multirow{2}{*} {$\delta_{\rm J2000}$}  & 
		\multicolumn{1}{c}{$R_{\rm proj}$}  &     
		\multicolumn{4}{c}{No. of candidate stars within...}\\ [0.5ex] 
		\cline{7-10}
		&  &  &  $hh$ : $mm$ : $ss$     & $^o$ : $^{\prime}$ : $^{\prime \prime}$ & kpc                & NW-K2 & M31 & MW   &Oth\\ [0.5ex] 
		\hline 
		NWS6       &  2013-09-12   &  D. Mackey  &  00:28:25.83   &  +40:45:54.74   &   37.6    &  11    &   18    & 27   &	                \\ [1.5ex]
		508HaS    &  2009-10-15   &  M. Rich       &  00:23:00.27   &  +41:12:45.91   &   50.8    &          &           &        & 77$^{(a)}$ \\ [1.5ex]			
		NWS5       &  2013-09-11   &  D. Mackey  &  00:20:03.15   &  +42:44:18.98   &   61.1    &    6    &   10    &  27  &                    \\ [1.5ex] 		
		507HaS    &  2009-10-15   &  M. Rich       &  00:17:58.08   &  +43:07:0.14     &   67.8    &    3    &    8     &  68  &                    \\ [1.5ex] 
		506HaS    &  2009-10-15   &  M. Rich       &  00:15:58.09   &  +43:58:48.47   &   77.0    &    3    &    7     &  72  &   1$^{(b)}$   \\ [1.5ex]		
		NWS3       &  2013-09-11   &  D. Mackey  &  00:13:33.45   &  +44:43:08.81   &   87.0    &    8    &    5     &  37  &   1$^{(b)}$    \\ [1.5ex] 
		505HaS    &  2009-10-15   &  M. Rich       &  00:12:55.57   &  +45:01:18.47   &   90.5    &    1    &    3     &   83  &  1$^{(b)}$    \\ [1.5ex] 		
		704HaS    &  2011-09-27   &  M. Rich       &  00:11:03.27   &  +45:32:44.0      &   98.1    &    4   &    5     &  80   &                      \\ [1.5ex]  
		606HaS    &  2010-09-09   &  M. Rich       &  00:08:36.35   &  +46:38:36.04   &  111.7    &    1   &    5     &  83   &   1$^{(b)}$    \\ [1.5ex] 	
		NWS1      &  2013-09-12   &  D. Mackey  &  00:07:56.18   &  +47:06:42.68    & 117.0    &     3   &    8     & 43   &   1$^{(b)}$      \\ [1.5ex] 
		\hline
	\end{tabular}
\end{table*}

\section{Observations} \label{Observations}
\graphicspath{ {Figures/} } 

Our photometric data were obtained within the Pan-Andromeda Archaeological Survey (PAndAS, \citealt{RefWorks:58}) using the 3.6m Canada-France-Hawaii Telescope (CFHT).  Equipped with the MegaPrime/MegaCam, that comprises 36, 2048 x 4612, CCDs with a pixel scale of 0.185 arcsec/pixel, it was able to deliver $\sim$1 degree$^2$ field of view \cite{RefWorks:58}.  \textit{g}-band (4140\AA{}- 5600\AA) and \textit{i}-band (7020\AA{}-8530\AA) filters were used to facilitate good colour discernment of Red Giant Branch (RGB) stars. Seeing of < 0.8 arcsec enabled stellar resolution to a depth of \textit{g} = 26.5 and \textit{i} = 25.5 with a signal to noise ratio of $\sim$10 (\citealt{RefWorks:58}, \citealt{RefWorks:42}, \citealt{RefWorks:81}).   

To determine the photometric zero points, de-bias, flat-field and fringe correct the data they were first processed by the Elixir system, \cite{RefWorks:570}, at CFHT. The data were then further reduced using a specifically constructed pipeline at the Cambridge Astronomical Survey Unit, \cite{RefWorks:336}.  Finally, morphological classifications (e.g. point source, non-stellar, noise) of the data were identified and stored along with \textit{g} and \textit{i} values on the PAndAS catalogue,  \cite{RefWorks:18}. For this work we select point source objects. 

\begin{figure*}
	\includegraphics[height=.24\paperheight, width=.86\paperwidth]{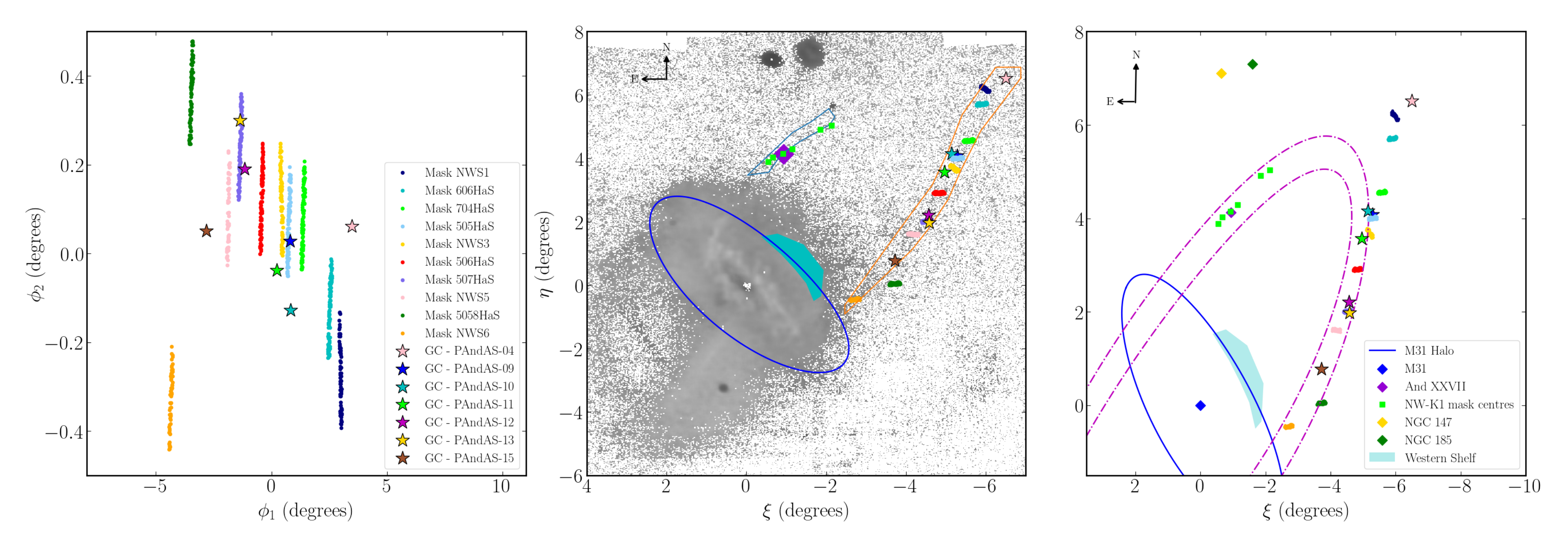}
    	\vspace*{-7mm}\caption[On sky locations of the masks]
	{On sky locations of the masks and GCs along NW-K2. The left hand panel shows the locations of all stars on each mask and the GCs plotted in stream coordinates.  M31 is not shown in this panel as it lies well below the left hand corner, out of range of both the x and y-axes, with coordinates of (-5.13 -2.97).  The middle panel shows NW-K1 (smaller open polygon) and NW-K2 (larger open polygon) superimposed over a stellar density map of M31. The right hand panel shows the locations of all stars on each mask and the GCs, plotted in tangent plane coordinates, along with other features, such as And XXVII.  The ellipse represents the M31 halo (taking a semi-major axis of  55 kpc with a flattening of 0.6, \citealt{RefWorks:107}). The dotted lines outline the inner and outer edges of an ellipse tracing the possible track of the NW Stream, assuming it to be a single feature (following the approach by \citealt{RefWorks:76}). The small circular icons represent observed stars, colour-coded to show their respective masks.	}
	\label{K2_Fig1}
\end{figure*}

Additional observations (see Table \ref{K2_table:1}) along the NW-K2 stream were taken with the Keck II telescope fitted with the DEep-Imaging Multi-Object Spectrograph (DEIMOS) using the OG550 filter with the 1200 lines/mm grating with a resolution of $\sim$1.1\AA {}-1.6\AA {} at FWHM. 
The masks were observed as follows: NWS3 and NWS6 were observed for a total of 1 hour 40 mins split into 5 x 20 minute integrations; 506HaS, 507HaS and 704HaS were observed for 1 hour 30 mins (3 x 30 minutes); NWS1 and NWS5 had a total observation time of 1 hour 20 minutes (4 x 20 minutes) and 606HaS had 3 x 15 minute integrations with a total observation time of 45 minutes.  

We selected our target stars based on their location within the colour magnitude diagram (CMD). Our highest priority targets were bright stars lying on the NW-K2 RGB with 20.3 < \textit{i$_0$} < 22.5, where \textit{i$_0$} is the de-reddened \textit{i}-band magnitude, given by \textit{i$_0$} = \textit{i} - 2.086E(B-V), obtained using extinction maps and correction coefficient, from Table 6, in \cite{M:1104}. Our next priority were fainter stars on the RGB, i.e. 22.5 < \textit{i$_0$} < 23.5. The remainder of the mask was filled with stars in the field with 20.5 < \textit{i$_0$} < 23.5 and 0.0 < \textit{g-i} < 4.0.  

The data from these observations were reduced using a pipeline, described in \cite{RefWorks:610}, which fitted resampled spectra with templates in the CaT region. The resulting cross-correlation functions were then fitted with Gaussians to determine the heliocentric velocity and uncertainty for each star.

Throughout this work we assume a radial velocity of $-$300 $\pm$ 4 {\kms} and an heliocentric distance of 783 $\pm$ 25 kpc for M31 (\citealt{RefWorks:56}).  With respect to this latter value we note that it precedes current findings for M31's heliocentric distance of 761 $\pm$ 11 kpc by \cite{M:899} and 798 $\pm$ 28 kpc by \cite{M:1091} but decide to continue using this value so our results are comparable to earlier works.  We also assume an heliocentric distance of 827 $\pm$47 kpc for And XXVII \cite{RefWorks:18}, \cite{RefWorks:586}.

\section{Analysis of NW-K2} \label{Analysing membership of NW-K2}
\subsection{Kinematics} \label{Kinematics}

To determine the properties of the NW-K2 stream we first identify and confirm members of its stellar population. As work by \cite{RefWorks:516} and \cite{RefWorks:405} noted a kinematic association between NW-K2 and the co-located GCs, we assume that NW-K2 stream stars will have velocities consistent with those of the GCs (see Table \ref{OM_table:2}) , i.e. between $\sim$ $-$397 \kms{ } for PAndAS-04 and $\sim$ $-$570 \kms{ }for PAndAS-13.  We therefore select stars from each mask with velocities in the range -600\kms to 0 \kms.  We also avoid any apparent failures in the pipeline and data with high uncertainties by refining our selection of stars to those with velocity uncertainties \mbox{< 20 \kms.}

We plot an initial set of velocity histograms for each mask and, taking the above criteria into account, find no NW-K2 candidate stars on mask 508HaS.  We note that this mask lies close to the edge of the stream (see Figure \ref{K2_Fig1}) and conclude that, as it was targeted on the same basis as the other masks, there is either a lower density of stream stars at this location or there is a gap in the stream.  This is consistent with findings by \cite{RefWorks:76} and \cite{RefWorks:405} who noted density variations along NW-K2.
\begin{table*}
	\centering
	\setlength\extrarowheight{2pt}
	\caption[Properties for NW-K2 globular clusters]
	{Properties for GCs, co-located on sky with NW-K2, from \cite{RefWorks:236}, \cite{RefWorks:233} and \cite{RefWorks:516}, with metallicities from \cite{M:992} .}		
	\label{OM_table:2}
	\begin{tabular}{lccccc} 
		\hline
		Name & $\alpha_{\rm J2000}$   &    $\delta_{\rm J2000}$   &   v$_r$   &   $R_{\rm proj}$  & [Fe/H] \\ [0.5ex]
		           & $hh$ : $mm$ : $ss$      & $^o$ : $^{\prime}$ : $^{\prime \prime}$  & {\kms}& kpc    &\\ [0.5ex]
		\hline 
		PAndAS-04  &  00:04:42.9    &   47:21:42.0     &   $-$397 $\pm$ 7    &  124.6  &  $-$2.07 $\pm$0.2  \\[0.5ex] 
		PAndAS-09  &  00:12:54.6    &   45:05:55.0     &   $-$444 $\pm$ 21  &    90.8  &  $-$1.56 $\pm$0.2  \\[0.5ex] 
		PAndAS-10  &  00:13:38.6    &   45:11:11.0     &   $-$435 $\pm$ 10  &    90.0  &                            \\[0.5ex] 
		PAndAS-11  &  00:14:55.6    &   44:37:16.0     &   $-$447 $\pm$ 13  &    83.2  &  $-$2.16 $\pm$0.2  \\[0.5ex] 
		PAndAS-12  &  00:17:40.0    &   43:18:39.0     &   $-$472 $\pm$ 5    &    69.2  &                            \\[0.5ex]  
		PAndAS-13  &  00:17:42.7    &   43:04:31.0     &   $-$570 $\pm$ 45  &    68.0  &                             \\[0.5ex]  
		PAndAS-15  &  00:22:44.0    &   41:56:14.0     &   $-$385 $\pm$ 6    &    51.9  &                             \\[0.5ex]  
		\hline
	\end{tabular}
\end{table*}

For the remaining masks we note that there are only a small number of potential NW-K2 stars on each and decide to proceed by combining the data from all masks for further analysis.  To enable us to obtain a consistent set of NW-K2 stars, determine any possible velocity gradient and create simple orbital models across the observed fields, we transform our data to an NW-K2 frame of reference. Using techniques described in, for example, \cite{RefWorks:421, RefWorks:537, M:1003} and \cite{M:1013, RefWorks:455, RefWorks:534} we convert the stellar coordinates from ($\alpha$, $\delta$) to NW-K2 centric coordinates ($\phi_1$, $\phi_2$) by rotating the celestial equator to a great circle where the pole, ($\alpha_{\rm pole}$, $\delta_{\rm pole}$) = (-64.01, -21.05) with a zero point azimuthal angle, $\phi_0$ = 138.58$^{\circ}$, in the new coordinates. Our rotation matrix is shown at Appendix \ref{Celestial coordinates}.  We also perform the same rotation on the co-located GCs.

To determine which stellar population (NW-K2, M31 or MW) a given star of velocity, $v_{\rm i}$ and velocity uncertainty of $v_{\rm err,i}$ is most likely to belong to, we define a single Gaussian function for each of them of the form:
\begin{equation} 
	\label{eq:3}
	\begin{multlined}
		P_{\rm struc}  = \frac{1}{\sqrt{ 2 \pi( \sigma_{v,\rm struc}^2 + v^2_{{\rm err},i} + \sigma_{\rm sys}^2)}}
		 \times \\  \\
		\shoveleft[1cm] \mathrm{exp}\Bigg[-\frac{1}{2} 
		\bigg( \frac{v_{r,\rm struc} - v_{r,i}} {\sqrt{\sigma_{\rm v,struc}^2 + v^2_{{\rm err},i} + \sigma_{\rm sys}^2)}}
		\bigg)^2 
		\Bigg],
	\end{multlined}
\end{equation}
where: $P$$_{\rm struc}$ is the resulting probability distribution function (pdf); $v$$_{\rm r, struc}$ ({\kms}) is the systemic velocity of the structure (i.e. NW-K2, M31 or MW); $\sigma$$_{\rm v, struc}$({\kms}) is the velocity dispersion of the structure,  $v$$_{\rm r, i}$ ({\kms}) is the velocity of each star on the masks, \mbox{$v$$_{\rm err, i}$ ({\kms})} the associated velocity uncertainty and $\sigma_{\rm sys}$ is a systematic uncertainty component of \mbox{2.2 {\kms}}, determined by \cite{RefWorks:293}, \cite{RefWorks:157}  and \cite{RefWorks:92} and applicable to our observations. The likelihood function for membership of NW-K2, based on velocity, is then defined as:
\begin{equation} 
	\label{eq:5}
	\begin{multlined}
		\mathrm{log}[\mathcal{L}(v_r, \sigma_r)] = \sum_{i=1}^{N} \mathrm{log}(\eta_{\rm M31} P_{i, \rm M31} +\\
		 \shoveleft[3cm]  \eta_{\rm MW} P_{i, \rm MW} + \eta_{\rm K2} P_{i, \rm K2}),
	\end{multlined}
\end{equation}
\\
where $\eta_{\rm M31}$, $\eta_{\rm MW}$ and $\eta_{\rm K2}$ are the fraction of stars within each stellar population, \textit{v}$_r$ includes \textit{v}$_{r \rm K2}$, \textit{v}$_{r \rm M31}$ and \textit{v}$_{r \rm MW}$ and $\sigma$$_{r}$ includes $\sigma$$_{r \rm K2}$, $\sigma$$_{r \rm M31}$ and $\sigma$$_{r \rm MW}$.

We incorporate the above equations, tailored for each stellar population into an Markoc Chain Monte Carlo (MCMC) analysis, using the {\sc emcee} software algorithm, \cite{RefWorks:571}, \cite{RefWorks:63} and set broad priors for each stellar populations on the masks i.e. :
\begin{itemize}
	\item systemic velocities are $-$600 $\le$ $v_{\rm K2}$ /{\kms} $\le$ $-$350,  $-$350 $\le$ $v_{\rm M31}$ /{\kms} $\le$ $-$290 and $-$170 $\le$ $v_{\rm MW}$ /{\kms} $\le$ 0.
	\item velocity dispersions are 0 $\le$ $\sigma$$_{v \rm K2}$ /{\kms} $\le$ 50 and 0 $\le$ $\sigma$$_{v \rm MW}$ /{\kms} $\le$ 150.  For M31 we know that the velocity dispersion changes with the on-sky projected distance, R, from the centre of the galaxy in accordance with Equation \ref{eq:41} from \cite {RefWorks:10} and \cite {RefWorks:235} (which we adopt for consistency with and comparison to our previous work in \citealt {RefWorks:586}, while recognising that there are more recent findings for the velocity dispersion across the M31 halo by \citealt {RefWorks:397}). So we calculate M31 velocity dispersions for each star based on the projected radius of their respective masks from the centre of M31and provide these as fixed values to our MCMC analysis. 
	\item the fraction parameters are 0 $\le$ $\eta$ $\le$ 1 with $\eta_{\rm K2}$ + $\eta_{\rm M31}$ + $\eta_{\rm MW}$  = 1.
\end{itemize}

\begin{equation} 
	\label{eq:41}
	\sigma_{v}(R) = \bigg(152 - 0.9 \frac{R}{1\: \rm{kpc}}\bigg)  \: \rm km s^{-1}  kpc^{-1},
\end{equation}

We set our Bayesian analysis to run for 100 walkers taking 100,000 steps with a burn-in of 50,000.  We use the {\sc emcee} algorithm to fit Gaussians and derive posterior distributions for the systemic velocity, velocity dispersion and fraction parameters for each stellar population.  To ensure that the chains have converged, we check the autocorrelation time ($\tau$) and find it to be in the range 50 < $\tau$ < 110.  This indicates that the number of steps is well above the recommended 10$\tau$ (\citealt{RefWorks:382}) and sufficient to deliver a robust number of independent samples and well constrained parameters. We also review the posteriors, see example at Appendix \ref{Corner plot}, to visualise the distribution and covariance of the various parameters.

Having obtained a Gaussian pdf for each of the three stellar populations, we derive the probabilities for each star belonging to a given population using:
\begin{equation} 
	\label{K2_eq:4}
	P_{\rm vel} = \frac{P_{\rm K2}}{P_{\rm M31} + P_{\rm MW} + P_{\rm K2}},
\end{equation}
\\
with the probability of being a contaminant given by: 
\begin{equation} 
	\label{K2_eq:5}
	P_{\rm contam} = \frac{P_{\rm M31} +  P_{\rm MW}} {P_{\rm M31} +  P_{\rm MW} + P_{\rm K2}},
\end{equation}

\begin{figure}
	\includegraphics[height=.3\paperheight, width=\columnwidth]{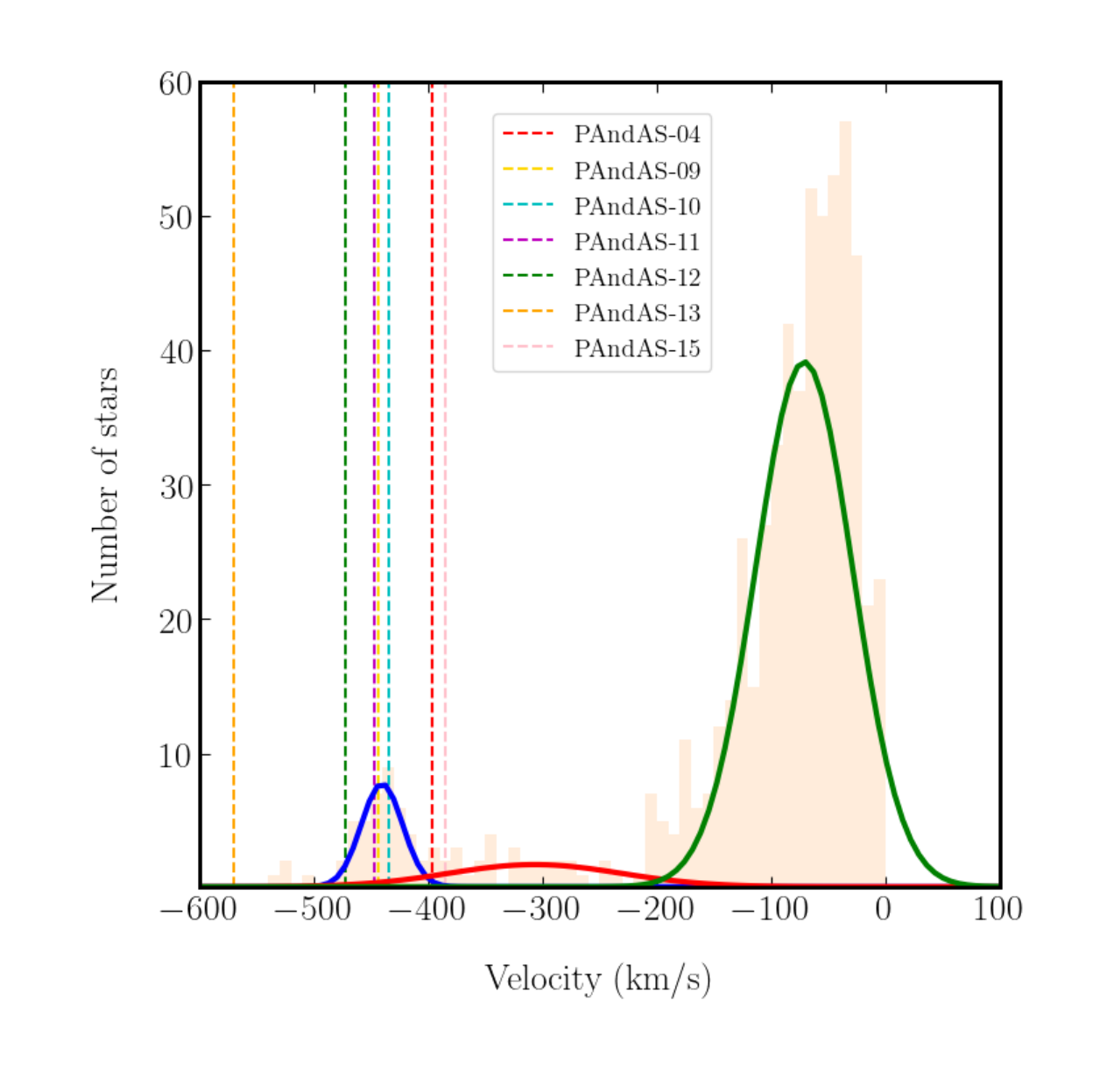}
    	\vspace*{-9mm}\caption[Velocity histogram for NW-K2 masks]
	{Kinematic analysis of NW-K2 fields showing the velocity histograms fields overlaid with the membership probability distribution function for each of stellar populations for, from left to right, NW-K2, M31 and the MW.  The plots also include the systemic velocities (vertical dotted lines) for the GCs co-located with NW-K2.
	}
	\label{K2_Fig2}
\end{figure}

We plot a velocity histogram, Figure \ref{K2_Fig2}, which reveals three kinematically distinct stellar populations: stars likely to be members of the MW (\textit{v$_r$} =  $\sim$ -80 {\kms}, \citealt{RefWorks:42}), stars likely to be members of the M31 halo (systemic velocity $\sim$ -300 {\kms}, \citealt{RefWorks:444}) and stars likely to be members of NW-K2 ($\sim$$-$600 $\le$ \textit{v$_{\rm K2}$}\kms $\sim$$\le$ $-$400). Also on this plot, we see clear indication that the GC, PAndAS-13, (orange vertical line) is unlikely to be associated with NW-K2, which is consistent with findings by \cite{RefWorks:236} and that PAndAS-15 (pink vertical line) and PAndAS-04 (red vertical line) are also on the periphery of association with the stream.

\begin{figure}
	\includegraphics[height=.3\paperheight, width=\columnwidth]{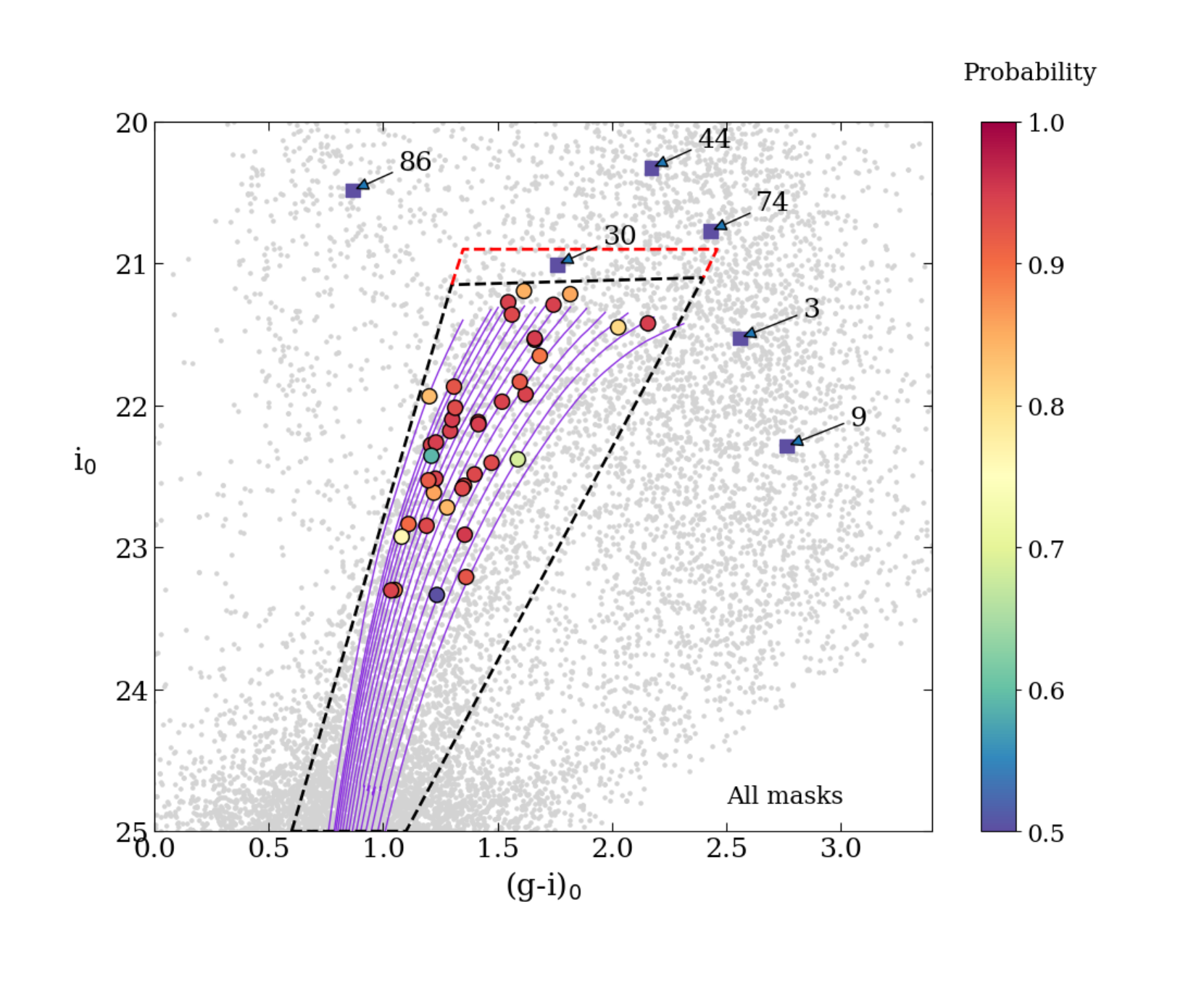}
    	\vspace*{-9mm}\caption[Isochrone analysis for masks]
	{CMD for NW-K2 candidate stars with an extinction and distance ({\Dsun} = 843 kpc) corrected array of isochrones aged 12 Gyrs,  [$\alpha$/Fe] = 0.0 and metallicities of $-$2.0 $\le$ [Fe/H] $\le$ $-$0.8 (moving from left to right across the plot) lying along the main spine of the RGB. The small dots show stars from the main PAndAS catalogue that lie within 15 arcmins of one of the masks (NWS6). The stars (circular icons) are colour coded by their strength of association with their nearest isochrone. The dashed line indicates the limits of the bounding box.  Stars outside the box are those with velocities similar to that of NW-K2 but which do not lie on the RGB and are, therefore, to be excluded from the NW-K2 stellar population. The additional, smaller bounding box indicates the extent of the TRGB using a distance correction of (m $-$ M) = 24.58 $\pm$ 0.19, which is the closest heliocentric distance reported for NW-K2 by \cite{RefWorks:405}.  
	}
	\label{K2_Fig4}
\end{figure}

We then refine the pool of possible NW-K2 stars by examining their proximity to the NW-K2 RGB.  Following an approach by \cite{RefWorks:38} and \cite{RefWorks:11}, we overlay the NW-K2 RGB with an array of isochrones whose metallicities cover the range of metallicities for the GCs i.e. $-$2.0 $\le$ [Fe/H] $\le$ $-$0.8 and that lie along the spine of the RGB.  We use the Dartmouth Stellar Evolution Database (\citealt{RefWorks:141}) to generate isochrones appropriate for the CFHT-MegaCam \textit{ugriz} filter, aged 12 Gyrs and with [$\alpha$/Fe]= 0.0 to form our array. We use the heliocentric distance of NW-K2 (843 $\pm$ 77 kpc, \citealt{RefWorks:405}) for the distance correction of the isochrones, which we also correct for reddening using E(B-V) = 0.08 as interpolated from the extinction maps in \cite{M:1104} by \cite{RefWorks:18}. We then plot the NW-K2 candidate stars (i.e. those with \textit{P$_{\rm vel}$} $\ge$ 50\%) onto the array of isochrones surrounded by a bounding box. Stars within the boundary of the box are very likely to be NW-K2 stars, but we cannot say definitively that they are.  However, we can have confidence that those lying outside the bounding box, further away from the NW-K2 RGB, are very unlikely to be members of NW-K2.  

We follow a technique used by \cite{RefWorks:92} to determine each star's probability of membership of NW-K2, and also to determine its photometric metallicity, based on its proximity to the nearest isochrone using:
\begin{equation} 
	\label{eq:1}
		P_{\rm iso}  = \rm exp\bigg(\frac{- \Delta (g-i)^2} {2 \sigma_c} - \frac{ \Delta (i)^2} {2 \sigma_m} \bigg),
\end{equation}
where $\Delta(g-i)$ and $\Delta(i)$ are distances from the isochrone nearest to the star, $\sigma$$_c$ is a free parameter that takes into account the range of colours of the stars on the CMD and $\sigma$$_m$ is a free parameter addressing distance and photometric errors. As this technique was used by \cite{RefWorks:586}, we adopt their values of $\sigma$$_c$ = 0.15 and $\sigma$$_m$ = 0.45 as our initial values and find that they deliver the appropriate results i.e. that stars lying well away from the NW-K2 RGB have a low probability of association with the isochrones. 

\begin{figure*}
  	\centering
	\includegraphics[height=.30\paperheight, width=.8\paperwidth]{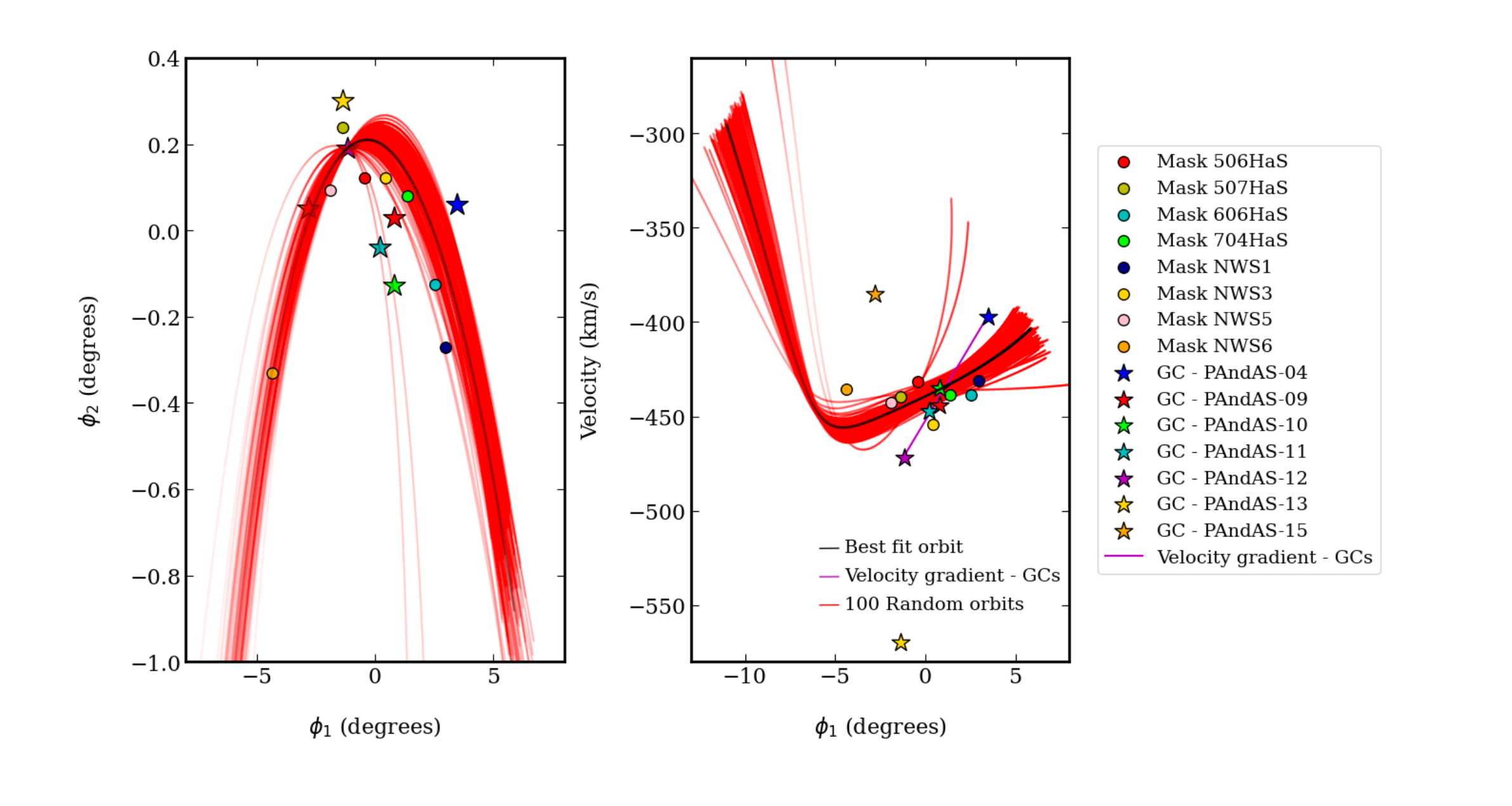}
    	\vspace*{-7mm}\caption[Orbital velocity of NW-K2]
	{The left hand plot shows the trajectory of NW-K2, in stream coordinates, created using a leapfrog integrator to generate orbits backward and forward from the stream's progenitor, which we assume to be the GC PAndAS-12.  The right hand plot shows the radial velocity of the stream relative to M31 together with the velocity gradient = \mbox{16$^{+3.2}_{-3.3}$ \kms degree$^{-1}$} across the GCs.  The velocity gradient along the stream through the observed fields is found to be \mbox{$-$1.2$^{+1.9}_{-1.8}$ \kms degree$^{-1}$} while the velocity gradient along the section of the orbit traversing these fields is determined to be 4.7 $\pm$ 0.004 \kms degree$^{-1}$. In both plots we show the data for 100 random orbits overplotted with the best fit orbit of the NW-K2 stream. 
		}
	\label{K2_Fig31}
\end{figure*}

\begin{figure*}
  	\centering
	\includegraphics[height=.3\paperheight, width=.6\paperwidth]{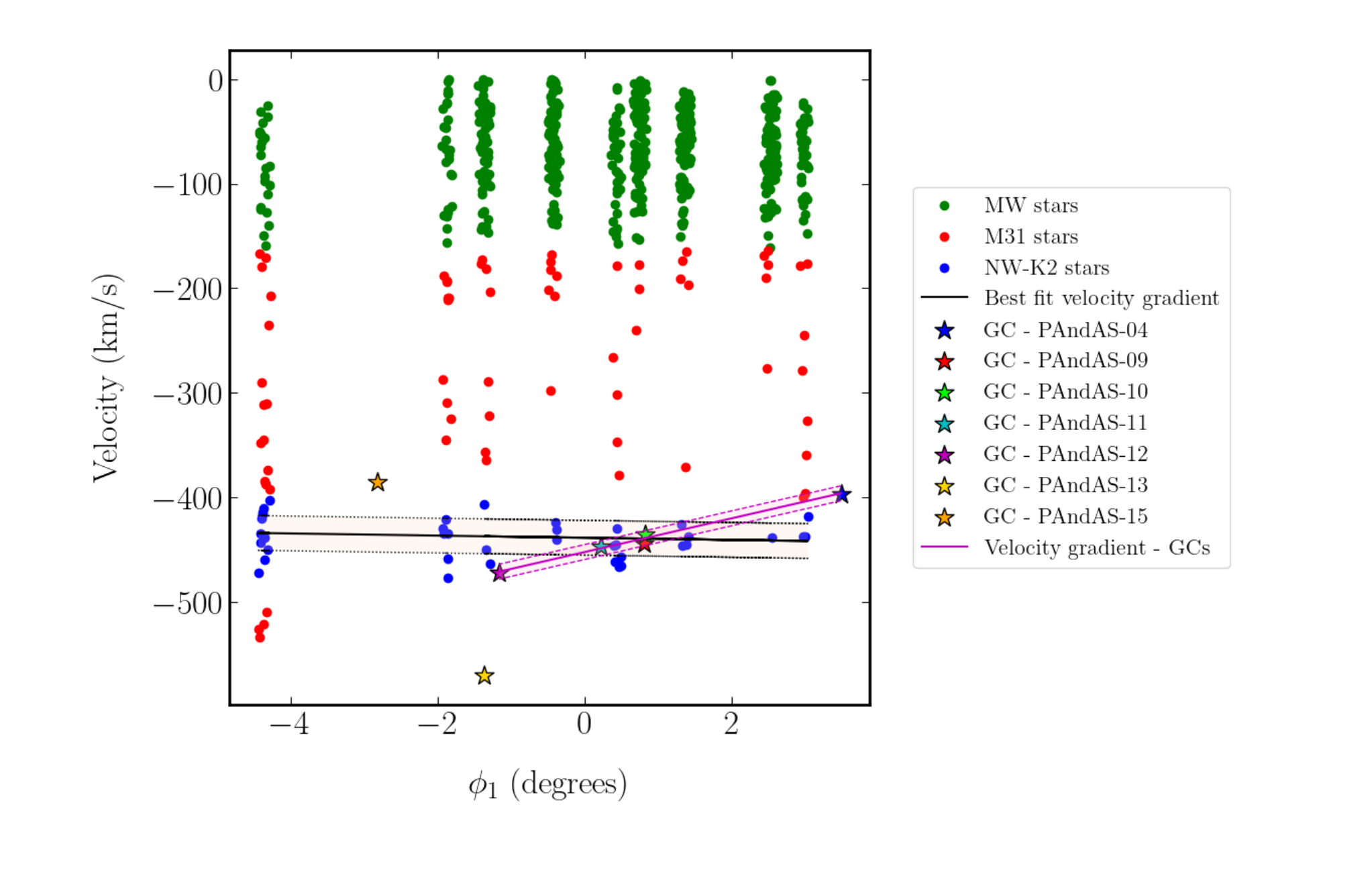}
    	\vspace*{-7mm}\caption[Velocity gradients across NW-K2 Stream fields]
	{Velocity gradient across NW-K2 obtained from our MCMC analysis of the observed fields.  The line has a slope of \mbox{$-$1.2$^{+1.9}_{-1.8}$\kms degree$^{-1}$} that is consistent with zero at 1$\sigma$.  The shaded area, bounded by dotted lines, indicates the velocity dispersion of the stream.  The velocity gradient of the GCs is also shown.
	}`
	\label{K2_Fig3}
\end{figure*}   

We note that star number 30 from mask 505HaS lies close to the top of the RGB just above the upper limit of our bounding box.  To see if this star could be a member of NW-K2, we re-run the isochrone analysis using using a distance modulus of 24.58 $\pm$ 0.19, this being the nearest heliocentric distance for NW-K2 reported by \cite{RefWorks:405}.  We denote the new location of the top of the bounding box with red dotted lines (see Figure \ref{K2_Fig4}) and see that star number 30 is now included as a possible member of NW-K2.

Our results also indicate that there are 5 candidate stars to be excluded from further analysis. These include:
\begin{itemize}
\item star number 74 on mask 505HaS.   As it lies close to the top of the bounding box it could be a very bright stream star.  The mask lies close to the GCs PAndAS-09 and PAndAS-10 but this star, with (g-i)$_0$ $\sim$ 2.4, is too red to be a member of either cluster where (g-i)$_0$ $\sim$ 0.62 and 0.75 respectively, \cite{RefWorks:233}. So it is more likely that this is an M31 halo star.
\item star number 9 on mask NWS3.   This star lies further away from the NW-K2 RGB so is unlikely to be a stream member. The mask lies close to PAndAS-11 but, again, with (g-i)$_0$ $\sim$ 2.8 the star is too red to be a member of this globular cluster where (g-i)$_0$ $\sim$ 0.67, \cite{RefWorks:233}. So it, too, is more likely to be an M31 halo star. 
\item star numbers 86, 44 and 3 on masks 506HaS, 606HaS NWS1, respectively, could be M31 halo stars. Integrating under the M31 gaussian we obtain an expectation $\sim$ 5-6 stars in the velocity range of all of our excluded stars (i.e. between $\sim$ $-$480 \kms and $\sim$ $-$400 \kms) so it is not implausible for all of them (and the above two stars) to belong to the M31 halo.  It is entirely possible that they were acquired inadvertently by the selection function that targeted RGB stars using colour selection boxes on the CMD, as described in \cite{RefWorks:669}, \cite{RefWorks:71} and \cite{RefWorks:18}.
\end{itemize}

We re-run our Bayesian analysis, this time excluding the stars that do not lie on the NW-K2 RGB.  We re-set our systemic velocity priors to $-$650 $\le$ $v_{\rm K2}$ /{\kms} $\le$ $-$370,  $-$450 $\le$ $v_{\rm M31}$ /{\kms} $\le$ $-$290 and $-$170 $\le$ $v_{\rm MW}$ /{\kms} $\le$ 0.  We retain the same values for the velocity dispersions and fraction parameters as described earlier.  To ascertain if there is a velocity gradient across the stream, we use techniques described by  \cite{RefWorks:257} and \cite{RefWorks:185}, and amend Equation \ref{eq:3} to include a velocity gradient ($\frac{dv}{d\phi_1}\ $) as shown in \ref{eq:9}.  We use the same number of walkers, steps and burn-in described previously and the same equations,  \ref{K2_eq:4} and \ref{K2_eq:5}, to determine probability of membership based on velocity. 
\begin{equation} 
	\label{eq:9}
	\begin{multlined}
		P_{\rm struc}  = \frac{1}{\sqrt{ 2 \pi( \sigma_{v,\rm struc}^2 + \sigma_{\rm sys}^2)}}
		 \times \\  \\
		\shoveleft[1cm] \mathrm{exp}\Bigg[-\frac{1}{2} 
		\bigg( \frac{\Delta v_{r,\rm i}} {\sqrt{\sigma_{v,\rm struc}^2 + \sigma_{\rm sys}^2)}}
		\bigg)^2 
		\Bigg],
	\end{multlined}
\end{equation}
where $\Delta v_{r,\rm i}$ ({\kms}) is the velocity difference between the \textit{i}$^{th}$ star and a velocity gradient, $\frac{dv}{d\phi_1}\ $ ({\kms}degree$^{-1}$) acting along the angular distance of the star's location on NW-K2, $\phi_1$, given by: 
\begin{equation} 
	\label{eq:10}
	\Delta v_{r,\rm i} = v_{r, \rm i}  - \Bigg( \langle v_{r} \rangle + \frac{dv}{d \phi_1} \phi_1 \Bigg),
\end{equation}

We then determine the overall probability each star's membership of NW-K2  by combining their probabilities of membership from the CMD and velocity analyses as follows:
\begin{equation} 
	\label{eq:101}
		 P_{\rm K2} = P_{\rm vel}  \times  P_{\rm iso},
\end{equation}

Our MCMC analysis returns a systemic velocity for NW-K2 = $-$439.3$^{+4.1}_{-3.8}$ \kms{} with a velocity dispersion = \mbox{16.4$^{+5.6}_{-3.8}$ \kms}. It also identifies an insignificant velocity gradient of\mbox{$-$1.2$^{+1.9}_{-1.8}$ \kms degree$^{-1}$}{} that is consistent with zero at 1$\sigma$.  This is not consistent with findings by \cite{RefWorks:236} who detected a velocity gradient along the stream, based on the properties of the GCs that they associated with it, of 1.0 $\pm$ 0.1\kms kpc$^{-1}$, which equates to 14.4 $\pm$ 1.4\kms degree$^{-1}$ in the same units as our gradient.

To explore this further, we plot the relative velocities of points along a number of model orbits for the NW-K2 stream.  The model orbits are generated following an approach described by \cite{RefWorks:455, RefWorks:534} that converts the data, $\alpha$, $\delta$ and velocity, to galactocentric values and then uses a leapfrog integrator to generate orbits backward and forward from the stream's progenitor, which, following the approach by \cite{RefWorks:313},  we assume to be the GC PAndAS-12.  The potential for M31 is modelled, as described by \cite{M:1148}, using parameters reported by  \cite{RefWorks:155, RefWorks:245}.  The properties of the model orbits are then converted to stream-centric values with the results, see Figure \ref{K2_Fig31}, showing a     similar disparity between their velocity gradients and that for the GCs. 

We plot the velocities of the stars in each stellar population as a function of position along the stream and show the results in Figure \ref{K2_Fig3}.  We then separate the data back into the component masks to determine the number of confirmed stars in each stellar population; to calculate a mean velocity for each mask and to undertake our spectroscopic analysis. We present the results of our kinematic analysis in Table \ref{K2_table:3}.

 \begin{table}
	\centering
	\setlength\extrarowheight{2pt}	
	\caption[Results of the kinematic analysis of NW-K2 fields]
		{Results of the kinematic analysis of NW-K2 fields. The table shows the mean velocity and the number of confirmed NW-K2 stars for each mask.}
		\label{K2_table:3}
	\begin{tabular}{lcc} 
		\hline
	 	Field & <v$_r$> & Confirmed\\
		 & {\kms}& stars\\
		\hline
NWS6     & $-$435.8       $\pm$ 19.8    &  11   \\[0.5ex] 
NWS5     & $-$442.5       $\pm$ 19.1    &   6    \\[0.5ex]  
507HaS   & $-$439.8      $\pm$ 24.1   &   3    \\[0.5ex]  
506HaS   & $-$431.5      $\pm$ 6.7     &   3   \\[0.5ex] 
NWS3     & $-$454.1      $\pm$ 11.9     &   8   \\[0.5ex]
505HaS   & $-$454.3     $\pm$ 0.0      &   1    \\[0.5ex]
704HaS   & $-$438.5     $\pm$ 8.3      &   4   \\[0.5ex]   
606HaS   & $-$438.41   $\pm$ 0.0      &   1   \\[0.5ex]   
NWS1     & $-$430.93    $\pm$ 9.1     &   3   \\[0.5ex] 
\hline
	\end{tabular}
\end{table}

\subsection{Metallicities}\label{Metallicities}

We measure the spectroscopic metallicities of the stars in NW-K2 using the CaT lines between 8400\AA-8700\AA.  As the relationship between equivalent width and [Fe/H] has been long established by \cite{RefWorks:318}, \cite{RefWorks:610} and \cite{RefWorks:272}, we fit a Gaussian function to the three CaT lines to obtain estimates of their equivalent widths and, following the approach by \cite{RefWorks:272} and \cite{RefWorks:42},  substitute our derived equivalent widths into:

\begin{equation} 
	\label{eq:21}
	\begin{multlined}
		[\mathrm{Fe/H]} = a + bM + cEW + dEW^{-1.5} + eEWM, 
	\end{multlined}
\end{equation}
\\
where: $a$, $b$, $c$, $d$ and $e$ are taken from the calibration to the Johnson-Cousins \textit{M$_I$} values and equal to -2.78, 0.193, 0.442, -0.834 and 0.0017 respectively; EW = 0.5EW$_{8498}$ +  EW$_{8542}$ + 0.6EW$_{8662}$ and  $M$ is the absolute magnitude of the star given by: 
\begin{equation} 
	\label{eq:22}
	M = i - 5 \times \mathrm{log}{_{10}}(\mathrm{D}{_{\odot}}) + 5,
\end{equation}	
where: $i$ is the $i$-magnitude of the star and $D{_{\odot}}$ is the heliocentric distance for the star, which, in keeping with the hypothesis that NW-K1 and NW-K2 are parts of a single structure, we assume to be the heliocentric distance for And XXVII.  The uncertainties on the equivalent widths are determined from the covariance matrix produced by the fitting process and these are combined in quadrature to yield the uncertainties on the metallicity.  

In some cases not all of the CaT lines are well resolved so we examine each spectrum by eye to determine which CaT lines are clearest for use in the determination of the metallicity. In the cases where we see too much noise around a CaT peak or one that has been affected by skylines, following the approach by \cite{RefWorks:42}, we ignore those lines and determine the value for EW using the more reliable lines, e.g. where the first CaT line is affected we take EW = EW$_{8542}$ + EW$_{8662}$, where only the second line is reliable we use EW = 1.7EW$_{8542}$ and where the third line is affected we assume EW = 1.5EW$_{8498}$ +  EW$_{8542}$.  

Having obtained metallicity values for the individual stars we then determine an average metallicity per mask and an average metallicity for NW-K2.  We show these results in the right hand column of Table \ref{table:2}.

As the S/N < 3 for many of the NW-K2 stars, we note that using the individual spectra to determine metallicity values may not deliver robust results. As the spread of metallicities on the CMD is not that large we stack the spectra (following the approaches by \citealt{RefWorks:444}, \citealt{RefWorks:160, RefWorks:175}, \citealt{RefWorks:52, RefWorks:171} and \citealt{RefWorks:586}) as combining them could provide a more accurate estimate of the mean metallicity for each mask.

\begin{figure}
	\includegraphics[height=.26\paperheight, width=\columnwidth]{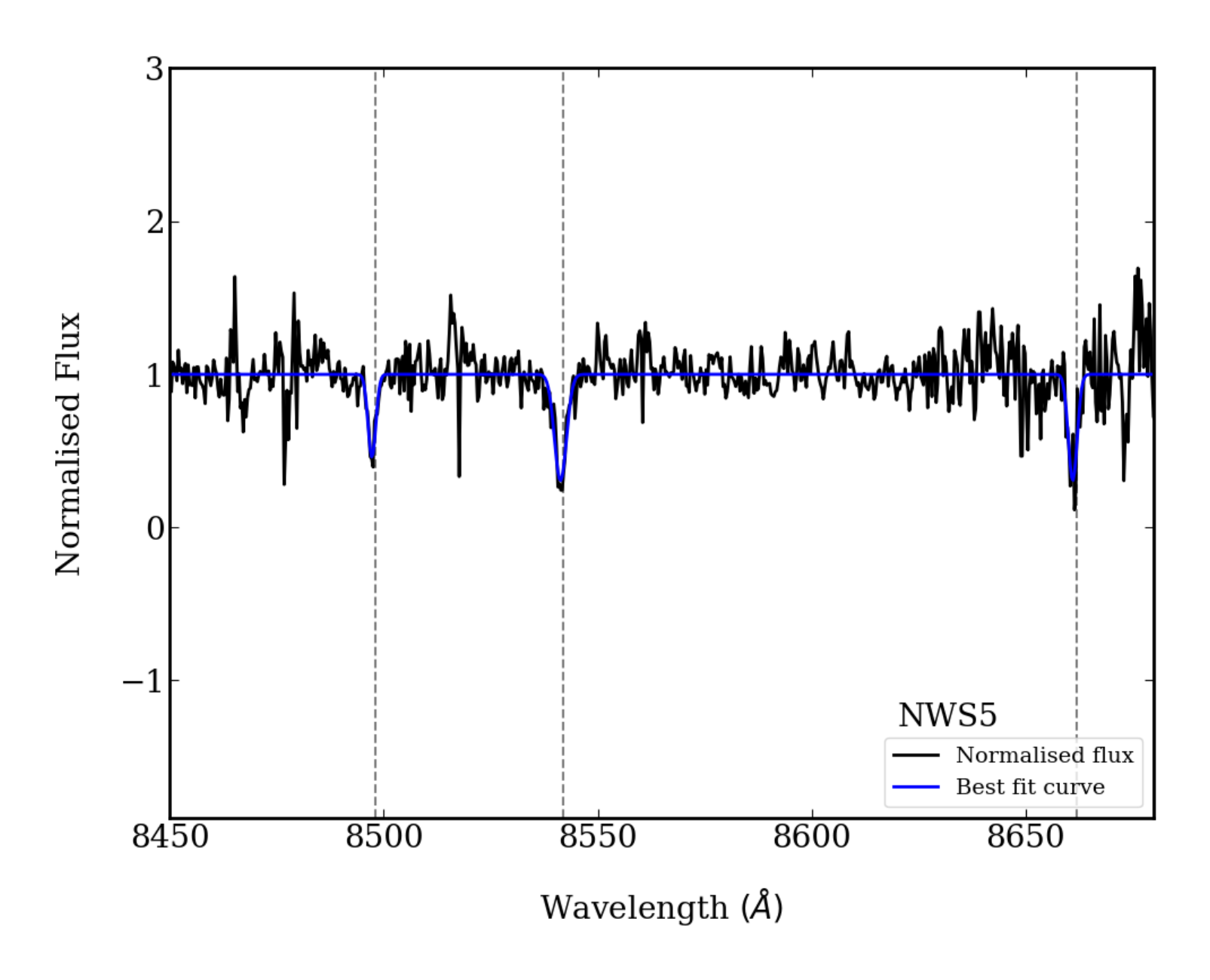}
    	\vspace*{-9mm}\caption[Example of co-added spectra for mask NWS5]
	{Example of co-added spectra for mask NWS5.  The normalised spectrum is overlaid with a best fit curve. The vertical dotted lines indicate the positions of the CaT lines.  This example is representative of the results for the other masks.
	}
	\label{K2_Fig5}
\end{figure}

We prepare the individual spectra following the approach described by \cite{RefWorks:42}, by correcting for the velocity of the individual star, smoothing the spectrum and normalising the data using a median filter.  We weight each spectrum by the S/N of its stars and interpolate to return the spectrum to the original lambda scale while retaining the velocity corrected position of CaT lines.  We derive a value for \textit{M} for the co-added spectrum by finding the average of the sum of the absolute magnitudes of the stream stars, each weighted by their S/N.  We simultaneously fit a Gaussian function to the CaT lines of the co-added spectra, as per the example shown at Figure \ref{K2_Fig5}, to obtain the equivalent widths and take the mean of the absolute magnitude values for the NW-K2 stars for use in the metallicity calculations described above.  Finally, we derive a mean metallicity for the stream and present our results in left hand column of Table \ref{table:2}.

\begin{table}
	\centering
	\setlength\extrarowheight{2pt}	
	\caption[Metallicities ]
	{Metallicities obtained from co-added spectra weighted by S/N for stars on each mask and an overall mean metallicity for the stream (left hand column) and from the average of the individual metallicities on the mask (right hand column).}		
	\label{table:2}
	\begin{tabular}{lcc} 
		\hline
		{Mask} & [Fe/H]$_{\rm spec}$ & <[Fe/H]$_{\rm spec}$>  \\ 
				\hline 
NWS6       &  $-$1.8$\pm$0.2  &  $-$1.4$\pm$0.2 \\[1.5ex]   
NWS5       &  $-$1.4$\pm$0.3  &  $-$1.2$\pm$0.1 \\[1.5ex]   
507HaS     &  $-$1.7$\pm$0.8  &  $-$1.4$\pm$0.4 \\[1.5ex]   
506HaS     &  $-$1.7$\pm$0.3  &  $-$1.5$\pm$0.2 \\[1.5ex]   
NWS3       &  $-$1.2$\pm$7.0  &  $-$1.4$\pm$0.2 \\[1.5ex]  
505HaS     &  $-$1.1$\pm$0.7  &  $-$1.1$\pm$0.7 \\[1.5ex]   
704HaS     &  $-$1.5$\pm$0.3  &  $-$1.3$\pm$0.1 \\[1.5ex]   
606HaS     &  $-$1.5$\pm$0.7  &  $-$1.5$\pm$0.7 \\[1.5ex]   
NWS1       &  $-$1.1$\pm$0.5  &  $-$1.2$\pm$0.1 \\[1.5ex]  

		\hline
NW-K2 mean &  $-$1.2$\pm$0.8  &  $-$1.3$\pm$0.1 \\[1.5ex]              
		\hline
	\end{tabular}
\end{table}

\section{Discussion} \label{Discussion} 

Our kinematic and spectroscopic analyses have confirmed a secure stellar population of 40 RGB stars for the NW-K2 segment of the NW stream.  We present our results in Table \ref{table:3}, with details of the properties of all the observed stars on the masks provided on-line in Appendix \ref{Properties of Observed Stars}.

\subsection{Kinematics}\label{Disc_Kinematics}

\begin{table*}
\vspace*{5mm}
\centering	
\setlength\extrarowheight{1pt}	
\vspace*{-5mm}\caption[Properties of the NW-K2 stream stars]
	{Table showing properties of NW-K2 candidate stars. The columns include: (1) mask name/star number; (2) right ascension in J2000; (3) declination in J2000; (4) \textit{i}-band magnitude; (5) \textit{g}-band magnitude; (6) signal to noise ratio; (7) line of sight heliocentric velocity; (8) metallicity. Photometric metallicity values, derived from proximity to a fiducial isochrone, are marked * where the spectrum was incomplete and ** where there was no spectrum for the star.  All other values are spectroscopic metallicities; (9) probability of membership of NW-K2 based on velocity, (10) probability of membership of NW-K2 based on proximity to a fiducial isochrone and (11) probability of membership of NW-K2 based on velocity and isochrone proximity. }
\label{table:3} 
\begin{tabular}{lcccccccccc}
\hline
Mask/star  & $\alpha$ & $\delta$ & \textit{i} & \textit{g} & \textit{S/N} &  \multicolumn{1}{c}{\textit{v$_r$}}   & [Fe/H] & {P$_{\rm vel}$} & {P$_{\rm cmd}$}& {P$_{\rm tot}$}   \\ 
                  &  $hh$ : $mm$ : $ss$     & $^o$ : $^{\prime}$ : $^{\prime \prime}$ &&&\AA$^{-1}$& \kms&&&&\\ [0.5ex] 
	\hline
NWS6 & & & & & &  & & &   \\ 
 2  & 00:28:54.13 & +40:43:32.5  &  22.5  &  23.7  &  2.3   & $-$472.3 $\pm$ 8.5  &   $-$1.0 $\pm$ 0.6   &  0.9   &  1.0   &  0.9 \\
 3  & 00:28:39.36 & +40:43:51.1  &  22.3  &  23.5  &  3.0   & $-$443.1 $\pm$ 6.2  &   $-$1.6 $\pm$ 0.7   &  0.9   &  1.0   &  0.9 \\
 8  & 00:28:08.56 & +40:44:44.8  &  21.4  &  23.6  &  0.9   & $-$459.8 $\pm$ 17.6 &   $-$1.6 $\pm$ 0.8   &  1.0   &  0.8   &  0.8 \\
21  & 00:28:19.34 & +40:45:54.7 &  23.3  &  24.3  &  1.0   & $-$438.2 $\pm$ 8.8  &   $-$1.2 $\pm$ 0.6   &  0.9   &  1.0   &  0.9 \\
24  & 00:28:37.95 & +40:46:11.6 &  22.9  &  24.0  &  1.7   & $-$410.4 $\pm$ 5.3  &   $-$1.2 $\pm$ 0.6   &  0.8   &  1.0   &  0.8 \\
26  & 00:28:43.14 & +40:46:35.3 &  22.7  &  24.0  &  2.3   & $-$416.0 $\pm$ 6.9  &   $-$1.1 $\pm$ 0.7   &  0.8   &  1.0   &  0.8 \\
39  & 00:28:04.05 & +40:48:29.8 &  22.4  &  23.6  &  2.3   & $-$402.7 $\pm$ 11.2  &   $-$1.6 $\pm$ 0.7   &  0.6   &  1.0   &  0.6  \\
42  & 00:28:31.57 & +40:44:32.9 &  21.9  &  23.1  &  4.2   & $-$420.3 $\pm$ 6.1  &   $-$1.5 $\pm$ 0.7   &  0.9   &  1.0   &  0.9 \\
46  & 00:27:50.43 & +40:45:04.9 &  21.3  &  22.8  &  6.4   & $-$450.0 $\pm$ 3.6  &   $-$1.4 $\pm$ 0.7   &  1.0   &  1.0   &  1.0 \\
56  & 00:28:41.35 & +40:46:01.6 &  21.5  &  23.5  &  7.5   & $-$413.9 $\pm$ 6.2  &   $-$1.4 $\pm$ 0.7   &  0.8   &  1.0   &  0.8 \\
58  & 00:29:01.52 & +40:46:07.2 &  21.4  &  22.9  &  5.0   & $-$434.3 $\pm$ 4.7  &   $-$1.7 $\pm$ 0.7   &  0.9   &  1.0   &  0.9 \\
		\hline
NWS5 & & & & & &  & & &  \\ 
 1   & 00:19:22.93 & +42:41:01.4  &  22.3  &  23.5  &  2.8   & $-$458.2 $\pm$ 9.0  &   $\sim$ $-$1.9* &  0.9   &  1.0   &  0.9 \\
 7  & 00:19:59.59 & +42:42:54.1 &  23.3  &  24.4  &  1.0  & $-$421.3 $\pm$ 6.1  &   $-$1.2 $\pm$ 0.7   &  0.9   &  1.0   &  0.9 \\
 8  & 00:19:42.04 & +42:43:04.9  &  22.8  &  24.0  &  1.9   & $-$434.7 $\pm$ 11.4 &   $-$1.1 $\pm$ 0.7   &  0.9   &  1.0   &  0.9 \\
16  & 00:19:48.46 & +42:44:11.3 &  22.8  &  23.9  &  1.8   & $-$477.0 $\pm$ 11.9 &   $-$1.0 $\pm$ 0.6   &  0.9   &  1.0   &  0.9 \\
20  & 00:20:40.26 & +42:44:38.1 &  21.2  &  23.0  &  6.7   & $-$429.5 $\pm$ 2.4  &   $-$1.4 $\pm$ 0.7   &  0.9   &  1.0   &  0.9\\
23  & 00:20:38.04 & +42:45:03.2  &  22.0  &  23.3  &  3.5   & $-$434.5 $\pm$ 6.4  &   $-$1.2 $\pm$ 0.7   &  0.9   &  1.0   &  0.9 \\
		\hline
507HaS & & & & & &  & & &  \\ 
17  & 00:18:09.16   &   +43:07:36.5  & 22.4  & 24.0  &  2.7   &  $-$406.7 $\pm$14.4     &$-$1.7 $\pm$ 0.8  &  0.7   &  0.9   &  0.7 \\
26  & 00:17:30.26 & +43:08:39.6 &  22.2  &  23.5  &  3.0   & $-$463.2 $\pm$ 14.9 & $-$0.9 $\pm$ 0.6    &  0.9   &  1.0   &  0.9 \\
49  & 00:17:29.70   & +43:04:51.0 &  22.1  &  23.4  &  3.7   & $-$449.6 $\pm$ 4.0  & $-$1.6 $\pm$ 0.7   &  0.9   &  1.0   &  0.9 \\
		\hline
506HaS & & & & & &  & & &  \\ 
46  & 00:15:39.35 & +44:00:00.9   &  21.9  &  23.2  &  4.9   & $-$430.2 $\pm$ 5.0  & $-$1.8 $\pm$ 0.7   &  0.9   &  1.0   &  0.9 \\
49  & 00:15:15.58 & +43:57:00.8   &  21.7  &  23.3  &  4.8   & $-$424.1 $\pm$ 3.7  & $-$1.4 $\pm$ 0.7   &  0.9   &  1.0   &  0.9 \\
50  & 00:15:25.92 & +43:58:55.9 &  22.0  &  23.5  &  4.5   & $-$440.4 $\pm$ 5.3  & $-$1.3 $\pm$ 0.7     &  0.9   &  1.0   &  0.9\\
		\hline
NWS3 & & & & & &  & & &  \\ 
 6  & 00:13:06.69 & +44:38:10.7  &  21.9  &  23.5  &  3.0   & $-$446.3 $\pm$ 5.4     &   $-$1.0 $\pm$ 0.6      &  1.0   &  1.0   &  1.0  \\
14  & 00:13:19.47 & +44:40:26.2 &  22.1  &  23.5  &  2.4   & $-$461.9 $\pm$ 8.1     &   $-$1.4 $\pm$ 0.8      &  0.9   &  1.0   &  0.9 \\
23  & 00:13:33.13 & +44:43:02.3  &  22.9  &  24.3  &  1.1   & $-$445.1 $\pm$ 10.4 &   $-$1.3 $\pm$ 0.7       &  0.9   &  1.0   &  0.9  \\
24  & 00:13:40.37 & +44:44:08.2  &  21.3  &  23.0  &  5.4   & $-$461.8 $\pm$ 5.9   &   $-$1.4 $\pm$ 0.7       &  0.9   &  1.0   &  0.9\\
28  & 00:13:43.67 & +44:45:27.5 &  21.8  &  23.4  &  3.5  & $-$429.6 $\pm$ 8.1    &   $-$1.5 $\pm$ 0.7       &  0.9   &  1.0   &  0.9 \\
29  & 00:13:32.96 & +44:45:41.8 &  22.4  &  23.9  &  1.8   & $-$466.6 $\pm$ 10.1  &   $-$1.2 $\pm$ 0.7      &  0.9   &  1.0   &  0.9 \\
33  & 00:13:35.29 & +44:47:28.1 &  23.2  &  24.6  &  0.9   & $-$456.4 $\pm$ 9.8    &   $\sim$ $-$0.8**        &  0.9   &  1.0   &  0.9 \\
43  & 00:13:55.01 & +44:49:41.3 &  22.6  &  23.9  &  1.9   & $-$465.0 $\pm$ 19.0  &   $-$1.7 $\pm$ 0.7      &  0.9   &  1.0   &  0.9 \\
		\hline
505Has & & & & & &  & & &  \\ 
30  & 00:13:20.68 & +45:00:48.9 &  21.0  &  22.8  &  9.1  & $-$454.3 $\pm$ 2.5   &   -1.1 $\pm$ 0.6   &  1.0   &  0.9   &  0.9  \\
		\hline
704HaS  & & & & & &  & & &  \\  
13  & 00:10:33.23 & +45:31:16.9 &  22.5  &  23.7  &  2.5   & $-$444.8 $\pm$ 5.2  &   $-$1.1 $\pm$ 0.6   &  0.9   &  1.0   &  0.9  \\
30  & 00:11:28.49 & +45:32:10.4 &  21.5  &  23.2  &  6.1   & $-$425.5 $\pm$ 3.0  &   $-$1.3 $\pm$ 0.7   &  0.9   &  1.0   &  0.9  \\
45  & 00:11:34.06 & +45:33:17.8 &  21.5  &  23.2  &  5.8   & $-$446.5 $\pm$ 3.5  &   $-$1.5 $\pm$ 0.7   &  0.9   &  1.0   &  0.9  \\
61  & 00:10:44.95 & +45:34:19.9 &  22.1  &  23.5  &  3.6   & $-$437.2 $\pm$ 4.3  &   $-$1.4 $\pm$ 0.7   &  0.9   &  1.0   &  0.9  \\
		\hline
606HaS  & & & & & &  & & &  \\  
42  & 00:08:14.30 & +46:37:57.9 &  21.2  &  22.8  &  5.4   & $-$438.4 $\pm$ 5.0  & $-$1.5 $\pm$ 0.7   &  0.9   &  1.0   &  0.9  \\
		\hline
NWS1  & & & & & &  & & &  \\ 
 6   & 00:07:31.97 & +47:03:08.5  &  22.6  &  23.9  &  1.9   & $-$437.1 $\pm$ 4.8  &   $-$1.1 $\pm$ 0.6   &  0.9   &  1.0   &  0.9\\
15  & 00:07:45.43 & +47:06:39.7 &  22.5  &  23.9  &  2.0   & $-$437.6 $\pm$ 8.9  &   $-$1.4 $\pm$ 0.7   &  0.9   &  1.0   &  0.9 \\
27  & 00:08:11.01 & +47:11:50.1 &  22.6  &  23.8  &  2.0   & $-$418.0 $\pm$ 11.7 &   $-$1.1 $\pm$ 0.6   &  0.9   &  1.0   &  0.9 \\

\end{tabular}
\end{table*} 

We find NW-K2 to have a systemic velocity of \mbox{$-$439.3$^{+4.1}_{-3.8}$ \kms} with a velocity dispersion of \mbox{16.4$^{+5.6}_{-3.8}$ \kms}. This is in keeping with the progenitor of NW-K2 being a dwarf galaxy and is consistent with other streams thought to have dwarf galaxy progenitors, see Table \ref{table:4}. With a current working assumption that the NW stream is a single structure, we also note these results are consistent with findings for NW-K1, \cite{RefWorks:586}.

We find plausible associations of the GCs PAndAS-04, PAndAS-09, PAndAS-10, PAndAS-11 and PAndAS-12 with NW-K2, which is in-keeping with findings by \cite{RefWorks:72, RefWorks:516},  \cite{RefWorks:236} and \cite{RefWorks:405}.  We also find that PAndAS-13 and PAndAS-15 (v$_r$ = $-$570 $\pm$ 45 \kms and \mbox{v$_r$ = $-$385 $\pm$ 6 \kms} respectively) are unlikely to be associated with NW-K2 despite their very clear co-locations. 
  
We obtain a velocity gradient of $-$1.2$^{+1.9}_{-1.8}$\kms degree$^{-1}$, that is consistent with zero at 1$\sigma$, along the stream with velocities becoming increasingly negative in the direction of M31.  This is not consistent with findings from \cite{RefWorks:236} who detected a stronger velocity gradient (across the GCs they associated with NW-K2) of 1.0 $\pm$ 0.1 \kms kpc$^{-1}$ (which equates to 14.4 $\pm$ 1.4\kms degree$^{-1}$ in the same units as our gradient) which they believed indicated that the stream progenitor was on an  infall trajectory towards M31. 

To explore the disparity in the velocity gradients, we fit the gradient model to only the GCs.  From this we obtain a systemic velocity of -452$^{+6.6}_{-6.4}$ \kms and a velocity dispersion of \mbox{7.1$^{+11.1}_{-5.1}$ \kms} which is indicative of a very cold system with the progenitor most likely to be a globular cluster (e.g: 300S, $\sigma_v$ $\sim$ 2.5 \kms and Ophicus, $\sigma_v$ $\sim$ 2.4 \kms, \citealt{M:924}). However, to have created a stellar stream containing, at least, 5 GCs, any progenitor of NW-K2 would have to be a significantly larger object with a much larger velocity dispersion and is, therefore, more likely to be a dwarf galaxy such as the dSph galaxies Sagittarius and Fornax, which have $\sigma_v$ $\sim$11.4 \kms and $\sim$11.8 \kms and host 8 and 5 GCs respectively, or NGC185 where  $\sigma_v$ $\sim$24 \kms and which hosts 8 GCs, \cite{RefWorks:493}. As a result, it is unlikely that the velocity gradient of \mbox{16$^{+3.2}_{-3.3}$ \kms degree$^{-1}$} that we find from this analysis (and which is consistent with that obtained by \citealt{RefWorks:236}) is representative of the velocity gradient of NW-K2.

\begin{table}
	\centering
	\setlength\extrarowheight{2pt}	
	\caption[Streams]
	{Streams likely to have dwarf galaxy progenitors.  Data for the table is sourced from (1) \cite{RefWorks:586}, (2) \cite{M:1073}, (3) \cite{RefWorks:11}, \cite{RefWorks:38}, (4) \cite{M:924} and (5) this work.
	}
		
	\label{table:4}
	\begin{tabular}{llll} 
		\hline
		Stream & Host & $\sigma_v$ (\kms) & [Fe/H]   \\ 
		\hline
NW-K1$^{(1)}$         &   M31  & 10.0 $\pm$ 4.0                             &  $-$1.4 $\pm$ 0.1                 \\[1.5ex]  
Sagittarius$^{(2)}$   &   MW   & 11.4 $\pm$ 0.7                             &  $-$1.2 < [Fe/H] <  $-$0.8      \\[1.5ex]
GSS$^{(3)}$             &   M31  & 11 $\leq$ $\sigma_v$ $\leq$ 33   &  $-$1.3 < [Fe/H] <  $-$0.4      \\[1.5ex]  
Palca$^{(4)}$            &   MW  & 13.4$^{+1.9}_{-1.4} $                    &  $-$2.02 $\pm$ 0.04               \\[1.5ex]  
Jhellum $^{(4)}$       &   MW  & 13.7$^{+1.2}_{-1.1} $                    &  $-$1.83 $\pm$ 0.05               \\[1.5ex]  
Elqui $^{(4)}$            &   MW  & 16.2$^{+2.3}_{-2.1} $                    &  $-$2.22 $\pm$ 0.06               \\[1.5ex] 
NW-K2$^{(5)}$         &   M31  & 16.4$^{+5.6}_{-5.8}$                    &  $-$1.3 < [Fe/H] <  $-$1.2        \\[1.5ex]  
Turranburra$^{(4)}$  &   MW  & 19.7$^{+3.9}_{-3.0}$                    &  $-$2.18$^{+0.13}_{-0.14} $    \\[1.5ex]               

	\end{tabular}
\end{table}

However, it is surprising that a stream of such scale has virtually no discernible velocity gradient, so we review our results, focusing on the data set for mask NWS6.  This mask has a number of NW-K2 candidate stars clustered around the overall systemic velocity for the stream, but it also has 4 outliers with velocities \mbox{$\sim$500 \kms} (see Table \ref{table:5}), which is significant given the total number of stars in our sample. Given the proximity of this mask to the M31 halo it is possible that these outliers are the "real" stream stars while the other candidate NW-K2 stars belong to the M31 halo.  These outliers have metallicities of -1.9 $\le$ [Fe/H] $\le$ -1.2 that are consistent with those of confirmed NW-K2 stars on other masks and they also lie on the NW-K2 RGB so it is possible that they are stream stars. However, our data analysis consistently identified them as non-NW-K2 stars. The only way to force them to be associated with the stream was to preferentially select them by tightening boundary conditions and defining rather than fitting data parameters (such as the percentage of stream stars in a given data set).  This did increase the systemic velocity on this mask, which in turn increased the velocity gradient to $\sim$5 \kms degree$^{-1}$, which is still much shallower than that for the GCs. While intriguing, this is not a robust result since it requires artificially tight priors.  For our results, we instead use broad priors for all masks. Figure \ref{K2_Fig3} shows the results of this analysis and indicates the stellar populations for NW-K2, M31 and the MW.  The leftmost line of the plot shows the results for NWS6 with the small group of outliers (in the bottom left hand corner) ostensibly colour coded as M31 halo stars.  

\begin{table}
	\centering
	\setlength\extrarowheight{2pt}	
	\caption[Streams]
	{Properties of the outlying stars on field NWS6 including line of sight heliocentric velocity, \textit v ({\kms}) and spectroscopic metallicity.  Star number 7, which has an incomplete spectrum, shows the photometric metallicity derived from its proximity to a fiducial isochrone.}	
	\label{table:5}
	\begin{tabular}{cll} 
		\hline
		Star no & v (\kms) & [Fe/H]   \\ 
		\hline
5      &  $-$533.92 $\pm$ 6.65       &    $-$1.2 $\pm$ 0.7   \\ 
7      &  $-$509.92 $\pm$ 12.70     &    $\sim$$-$1.9         \\
9      &  $-$526.32 $\pm$ 3.64       &    $-$1.3 $\pm$ 0.7   \\
49    &  $-$520.96 $\pm$ 2.73       &    $-$1.7 $\pm$ 0.7   \\
	\end{tabular}
\end{table}

\begin{figure}
	\includegraphics[height=.26\paperheight, width=\columnwidth]{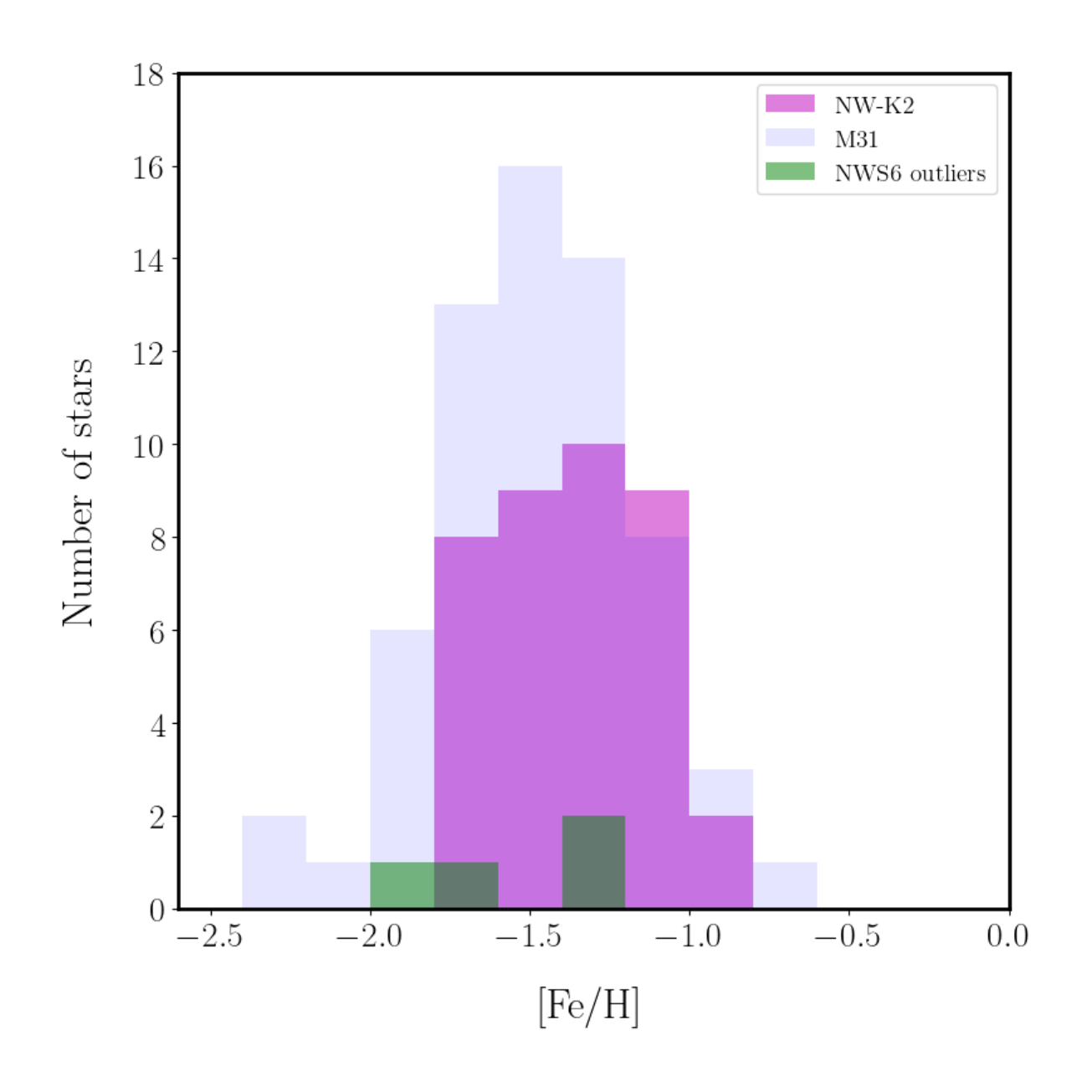}
    	\vspace*{-9mm}\caption[Metalllicity distributions]
	{Metallicity distributions of M31, NW-K2 and the NWS6 outlier stars.
	}
	\label{K2_Fig6}
\end{figure}

So what do these 4 stars represent?  The possibilities include:
\begin{enumerate}
  \item Despite the fitting process excluding these outliers from NW-K2, given their proximity to M31 and the consistency of their spectroscopic metallicities with those of NW-K2, see Figure \ref{K2_Fig6}, they could be high velocity stream stars and members of that stellar population.  
 \item Extrapolating from Figures 6 and 7 from \cite{M:1067} and Figure 5 from \cite{M:977}, none of which extend as far out as NWS6, it is possible that these outliers are part of M31's outer halo and could be members of the M31's thin or thick disk stellar populations.  \cite{RefWorks:171} produced a contour map of expected velocities of stars in circular orbits around M31 which, at a similar projected radius are lower than those of the outliers.  \citeauthor{RefWorks:171} also report a velocity dispersion of 50.8 $\pm$ 1.9 \kms and 35.7 $\pm$ 1.0 \kms for the thick disk and the thin disk respectively both of which of which are higher than the velocity dispersion across the outliers, where $\sigma$$_v$ $\sim$ 9 \kms.  This velocity dispersion is also inconsistent with velocity dispersion profiles reported by \cite{RefWorks:397} at a similar projected radius in the M31 halo, making it unlikely that the outliers are members of this stellar population.
  \item They could be Western Shelf stars.  Simulations of the Western Shelf, considered to have formed during the merger of M31 with a progenitor that created the Giant Stellar Stream in the south west of M31's halo, show a broad diffused structure to its outer edge, see \cite{RefWorks:155}, \cite{RefWorks:109} and \cite{M:1086}, that could be intersecting NW-K2.  However, \cite{RefWorks:109} find a metallicity distribution for the Western Shelf of $-$0.8 $\le$ [Fe/H] $-$0.5, which is more metal rich than the 4 outliers, and heliocentric velocities greater than those of the outliers at a similar projected radius, so it is unlikely that these outliers are members of the Western Shelf.
  \item Given their low velocity dispersion, they could be the cold component of a previously undiscovered substructure within the M31 halo. 
\end{enumerate}
This leaves us with an intriguing conundrum as to the nature of the 4 outliers, with the only way to determine what they really do represent being to obtain additional data, e.g. perhaps by extension of DESI coverage, as suggested by \cite{M:977}, and/or by the Subaru Prime Focus Spectrograph which is in the final stages of development and scheduled to become a Subaru Strategic Programme from 2024 \cite{M:1134}, \cite{M:1135}.

With no discernible gradient along NW-K2, it is unwise to postulate whether it and NW-K1 form a single structure or are separate streams.  However, as these gradients are based on line of sight velocities we recognise that there could be stronger, undetected, velocity components acting along the streams in other directions and that could affect their overall directions of motion.  Further modelling of the 3-d trajectories of both segments of the NW stream will be required ascertain the structure of the NW stream (\citealt{M:1148}).

\subsection{Metallicities}\label{Disc_Metallicities}

We determine $-$1.3 $\pm$ 0.1 $\le$ <[Fe/H]$_{\rm {spec}}$> $\le$ $-$1.2 $\pm$ 0.8 for the stream, with most stars in the NW-K2 stellar population having \mbox{$-$1.8 <[Fe/H]$_{\rm spec}$ < $-$0.9}, which is consistent with work by \cite{RefWorks:82}.  This is also consistent with findings of the metallicities of many Local Group dwarf galaxies \cite{RefWorks:56} who used the mass-metallicity relationship to estimate that the stellar mass of the NW stream progenitor could be $\sim$ 10$^{6-8}$ \Msun, indicative of it being a dwarf galaxy.  However, we do not find a metallicity gradient along the stream which is somewhat surprising since most dwarf galaxies have been found to have a range of metallicities that become increasingly metal-poor with distance from their centres (\citealt{M:842}) so we would expect to see a similar trend in their stellar streams.  However, \citeauthor{M:842} also found flatter gradients in galaxies with younger stellar populations (0-5 Gyrs) which aligns with findings by \cite{RefWorks:654} who classified the NW-K2 GCs in a sub-group of GCs accreted by M31 more recently (e.g. 2-3 billion years) than the other M31 GCs. \cite{M:1045} found flatter metallicity gradients in dwarf irregular and late-type spiral galaxies, so it is possible that one of these galaxies, or a dwarf galaxy with a very small innate metallicity gradient, is the progenitor of NW-K2.

\section{Conclusions} \label{Conclusions} 
In this work we present the results of our kinematic and spectroscopic analyses of 40 RGB stars from 9 fields spanning the length of the NW-K2 segment of the NW Stream.  We have identified secure members of the NW-K2 stellar population based on significant similarities in velocities and strong ($\ge$1$\sigma$) association with a grid of fiducial isochrones.

Our results do not conclusively indicate whether or not NW-K2 and NW-K1 are elements of the same, single, structure.  If they are, then it is very likely that And XXVII is the progenitor of both of them.   However, if they are separate streams then, while it is generally accepted that And XXVII is the progenitor of NW-K1, the progenitor of NW-K2 is, at present, undetected.  Our findings indicate that this progenitor could be a dwarf irregular or late-type spiral galaxy, accreted by M31 within the last $\sim$ 5 Gyrs.  Other possibilities are that the progenitor could be:
\begin{enumerate}
  \item a more massive system that collided with M31 on a radial orbit creating some of the other substructures (e.g. The East Cloud and South West Cloud) in the M31 halo.  Such an event would have expelled significant volumes of stars and GCs out to large radii and could have created tangential streams, such as NW-K2.  This would also provide a plausible explanation as to how this faint feature is associated with so many GCs, \cite{RefWorks:449}.
 \item an ultra diffuse galaxy (UDG), some of which have been found to host significant GC system, \cite{M:1098}, \cite{M:1099},  \cite{M:1095}, \cite{M:1097} and \cite{M:1096}.  Indeed \citeauthor{M:1097}, in their analysis of UDGs in the Coma system, postulate that UDGs with large GC abundances may have formed their GCs and fields stars early on and very quickly, giving rise to metal-poor stellar populations.  Failure of these UDGs to undergo subsequent star formation would yield the high specific frequencies (i.e. the number of GCs per unit galaxy luminosity) associated with these galaxies.
\end{enumerate}
 
NW-K2 stretches out 100 kpc, and possibly beyond, to the north west of M31’s halo.  It is indeed an intriguing feature that could yet provide more insights into its own formation and that of M31. Further detections of the stream, both to the southeast and northwest of M31, might locate its progenitor and yield the opportunity to better constrain the shape and size of the M31 potential. 

\section{Acknowledgements}

The authors wish to thank the anonymous reviewer for their insightful comments and advice. JP wishes to thank Stuart Sullivan, Joan Sullivan and Barry Sullivan for their inspiration.  JP also wishes to thank Mark Fardal and Dougal Mackey for their generous donations of data and private communications.

This work used the community-developed software packages: Matplotlib (\citealt{RefWorks:616}), NumPy (\citealt{RefWorks:615}) and Astropy (\citealt{RefWorks:613, RefWorks:614, M:1115}).

Most of the observed data presented herein were obtained at the W.M. Keck Observatory, which is operated as a scientific partnership among the California Institute of Technology, the University of California and the National Aeronautics and Space Administration. The Observatory was made possible by the generous financial support of the W.M. Keck Foundation. Data were also used from observations obtained with MegaPrime/MegaCam, a joint project of CFHT and CEA/DAPNIA, at the Canada-France-Hawaii Telescope which is operated by the National Research Council of Canada, the Institut National des Sciences de l'Univers of the Centre National de la Recherche Scientifique of France, and the University of Hawaii. The authors wish to recognise and acknowledge the very significant cultural role and reverence that the summit of Mauna Kea has always had within the indigenous Hawaiian community. 
\\
\section{Data Availability}
The data used in this paper are available herein, in \cite{RefWorks:586} and in their associated on-line supplementary materials. The raw DEIMOS data are available via the Keck archive.\\


\bibliography{BiblogNWK2}
\bibliographystyle{mnras}

\newpage
\appendix

\onecolumn   

 \section{Celestial coordinates ($\alpha$, $\delta$) to stream coordinates ($\phi_1$, $\phi_2$) transformation}  \label{Celestial coordinates}

\begin{equation}
\label{eqk27}
\begin{pmatrix}
	\rm cos(\phi_1)\rm cos(\phi_2) \\
	\rm sin(\phi_1)\rm cos(\phi_2) \\
	\rm sin(\phi_2)  
\end{pmatrix} 
=
\\[1ex]
\begin{pmatrix}
	$-$0.0548755604 & +0.4941094279    &  $-$0.8676661490 \\
	$-$0.8734370902 & $-$0.4448296300 &  $-$0.1980763734  \\
	$-$0.4838350155 & +0.7469822445    & +0.4559837762
\end{pmatrix}
\rm  \cdot  
\\[1ex]
\begin{pmatrix}
	\rm cos(\alpha)\rm cos(\delta) \\
	\rm sin(\alpha)\rm cos(\delta) \\
	\rm sin(\delta)  
\end{pmatrix}
\end{equation}

\clearpage

\section{Posteriors for the stellar populations of NW-K2, M31 and MW} \label{Corner plot}

\begin{figure*}[htp]
	\includegraphics[height=.65\paperheight, width=.75\paperwidth]{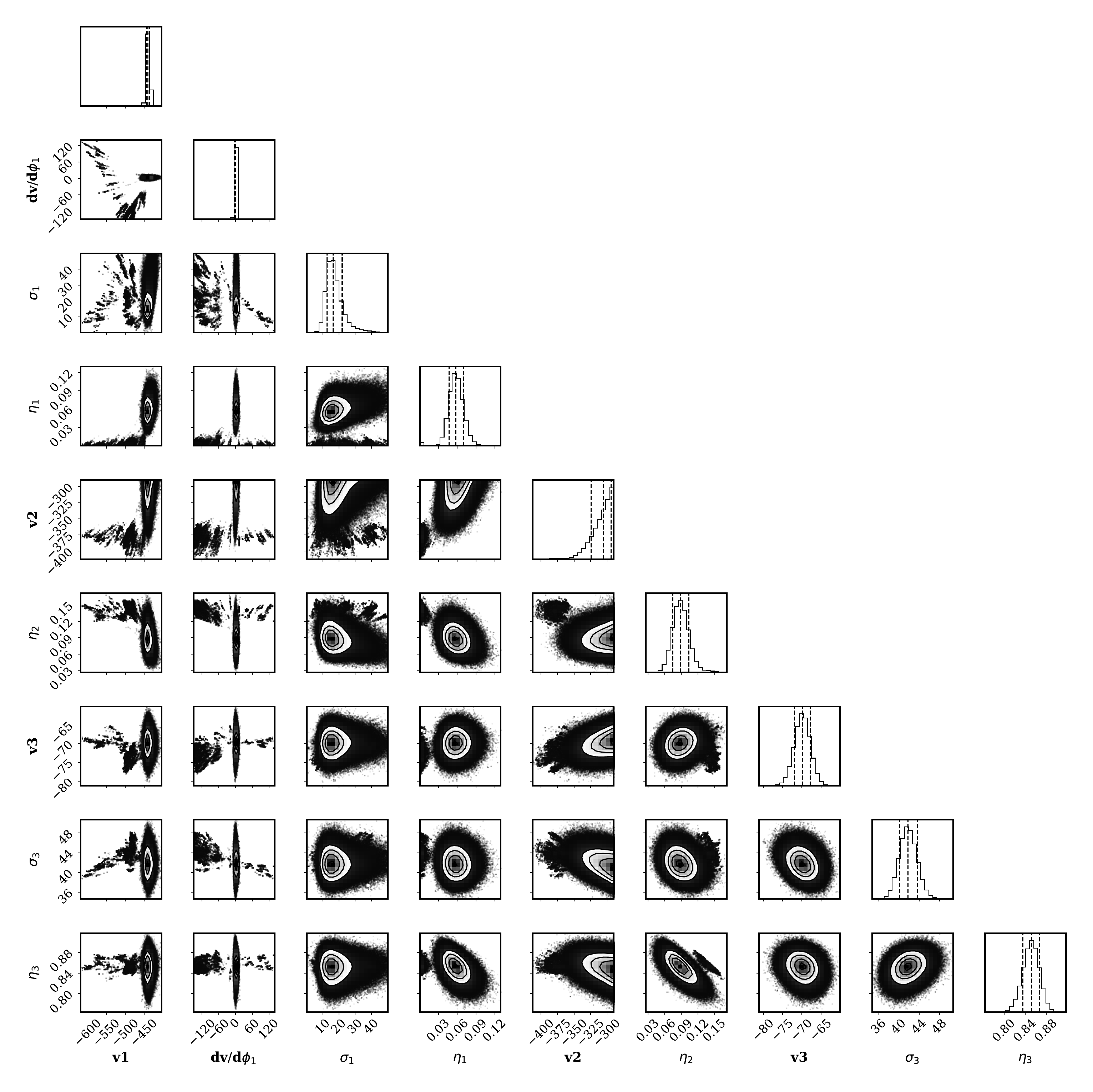}
    	\vspace*{-3mm}\caption
	{Posteriors for the stellar populations of NW-K2, M31 and MW where: 
	\textit{v}$_{\rm 1}$, \textit{v}$_{\rm 2}$ and \textit{v}$_{\rm 3}$ are the systemic velocities of each stellar population (NW-K2, M31 and MW respectively);  
	\textit{s}$_{\rm 1}$ and \textit{s}$_{\rm 3}$ are the velocity dispersions for NW-K2 and MW respectively (NB: the velocity dispersion for M31 was not fitted as it varied with the distance of each star from the centre of M31); 
	$\eta_{\rm1}$, $\eta_{\rm 2}$ and $\eta_{\rm 3}$ are the fraction of stars within the NW-K2, M31 and MW stellar populations respectively 
	and $\frac{dv}{d\phi_{1}}\ $ is the velocity gradient along the stream. 
	In each of the diagonal posteriors we indicate the median value for each parameter together with the 1-$\sigma$ uncertainties. 
	}
	\label{K2_Fig16}
\end{figure*}

\section{Properties of Observed Stars from the masks} \label{Properties of Observed Stars}

Appendix  \ref{Properties of Observed Stars} is available as a separate on-line document.

\end{document}